%% file: BoundEffRed.tex
\documentclass[journal]{IEEEtran}

\usepackage{amsmath,amssymb,amsfonts,amsthm}
\usepackage{euscript}
\usepackage[latin1]{inputenc}
\usepackage[T1]{fontenc}
\usepackage{color}
\usepackage{graphicx}
\usepackage{mathrsfs}
\usepackage{algorithm}
\usepackage{algorithmic}
\usepackage{dsfont}
\usepackage[shortlabels]{enumitem}
\usepackage{multirow}
\usepackage{hhline}
\usepackage{hyperref}
\usepackage{mathpazo}
\usepackage{array}
\usepackage{bigints}

\input{mydefs}

\title{An Efficient Forecasting Approach\\ to Reduce Boundary Effects in Real-Time Time-Frequency Analysis}

\author{Adrien~Meynard, %~\IEEEmembership{Member,~IEEE,}
        Hau-Tieng~Wu
\thanks{A. Meynard is with the Department
of Mathematics, Duke University, Durham,
NC, 27708 USA.

H.-T. Wu is with the Department of Mathematics and Department of Statistical Science, Duke University, Durham, NC, 27708 USA; Mathematics Division, National Center for Theoretical Sciences, Taipei, Taiwan.
 
A. Meynard is the corresponding author (e-mail: adrien.meynard@duke.edu).
}}

\newtheorem{remark}{Remark}

\newtheorem{theorem}{Theorem}

\begin{document}

\maketitle

\begin{abstract}
Time-frequency (TF) representations of time series are intrinsically subject to the boundary effects. As a result, the structures of signals that are highlighted by the representations are garbled when approaching the boundaries of the TF domain. In this paper, for the purpose of real-time TF information acquisition of nonstationary oscillatory time series, we propose a numerically efficient approach for the reduction of such boundary effects. The solution relies on an extension of the analyzed signal obtained by a forecasting technique. In the case of the study of a class of locally oscillating signals, we provide a theoretical guarantee of the performance of our approach. Following a numerical verification of the algorithmic performance of our approach, we validate it by implementing it on biomedical signals.
\end{abstract}


\section{Introduction}
\label{se:introduction}
\IEEEPARstart{I}{n} any digital acquisition system, the study and the interpretation of the measured signals generally require an analysis tool, which enables researchers to point out the useful characteristics of the signal. The need for signal analysis arises from various signals, ranging from audio~\cite{Stowell18computational,Muller11signal}, mechanical~\cite{Peng02vibration}, or biomedical signals~\cite{Akay96detection}. For instance, biomedical signals, such as photoplethysmogram (PPG), contain several characteristics, including respiratory rate or blood pressure, that cannot be interpreted from its run-sequence plot in the time domain. An analysis tool would make possible the extraction of these useful characteristics. 

Usually, the measured signals exhibit nonstationary behavior, and the observed quantities might be interfered with by transient phenomena that can vary rapidly and irregularly. In this paper, we focus on oscillatory time series. These signals might, for example, oscillate fast with large amplitudes at one moment, and then oscillate slowly with small amplitudes at the next moment. In order to adapt the analysis to nonstationarities, local spectral analysis is generally performed~\cite{Stoica05spectral,Matz97generalized}. The short-time Fourier transform~\cite{Grochenig01foundations} (STFT), a typical tool built for this purpose, enables the determination of the local frequency content of a nonstationary signal.

\begin{figure}
\centering
\includegraphics[width=.48\textwidth]{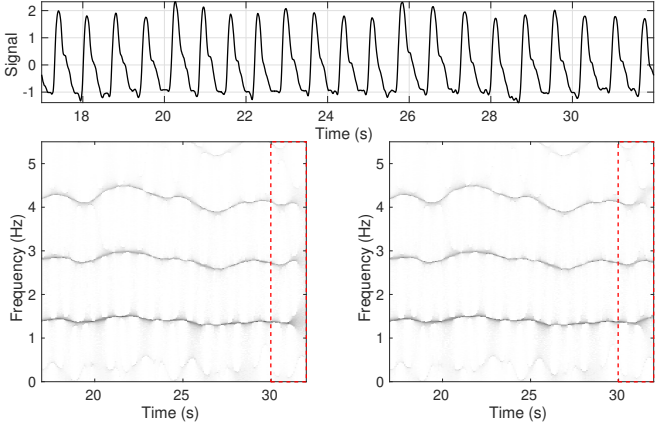}
\caption{A segment of PPG signal (top) and the right boundary of a TF representation determined by the SST without extension (bottom left), and the right boundary of a TF representation determined by the SST with the proposed boundary effect reduction algorithm by forecasting (bottom right). The window length for the SST is 12~seconds. Videos of these real-time SSTs are available online at \url{https://github.com/AdMeynard/BoundaryEffectsReduction}.}
\label{fig:ex.intro}
\end{figure}

Windowing is a common method for performing local analysis. Among many others, STFT~\cite{Flandrin:1999}, continuous wavelet transform (CWT)~\cite{Da1992}, synchrosqueezing transform (SST) \cite{Daubechies11synchrosqueezed}, and reassignment~\cite{Auger13time} (RS) are representations that fall back on the use of an analysis window. Let $x:I\to\RR$ denote the observed signal, where $I$ denotes the finite interval where the signal is measured. Let $g_s:\RR\to\RR$ denote the analysis window, where $s$ is a shape parameter. The support of $g_s$ is localized around the origin and is small with respect to $|I|$. The translation operator is $T_\tau$ defined as:
\[
T_\tau f = f(t-\tau)\ ,\quad \forall\ f:\RR\to\RR\ .
\]
Then, the local analysis of $x$ around the instant $\tau\in I$ relies on the evaluation of the following inner product:
\begin{equation}
V_x(s,\tau) = \langle x, T_\tau g_s \rangle_I \ .
\label{eq:windowing}
\end{equation}
A major shortcoming of this technique occurs when analyzing the signal $x$ {\em near the boundaries} of the interval $I$. Clearly, at these points, half of the information is missing. Consequently, the results of the inner product~\eqref{eq:windowing} are distorted. This phenomenon is usually understood as the \emph{boundary effect}. The result of the SST of a PPG is shown in the lower-left corner of Fig.~\ref{fig:ex.intro} (see Section~\ref{ssse:ppg} for a comprehensive description). The distortion resulting from boundary effects is clearly visible on the right side of this representation---this area is highlighted by the red dashed rectangle. Indeed, while in the major left part of the image, clear lines stand out, they become blurred as they approach the right boundary of the image. Therefore, estimations of signal characteristics, like instantaneous frequencies~\cite{Delprat92asymptotic} or amplitudes, from this TF representation appear to be imprecisely determinable (or even likely to fail) in the vicinity of the boundaries. Moreover, this boundary effect would unavoidably limit applying these window-based TF analysis tools for real-time analysis purposes. It is thus desirable to have a solution to eliminating the boundary effects.

Attempts to minimize the boundary effects generally consists in softening the discontinuity on signal edges. Usually, the approaches can be classified into two main classes, choosing a proper analysis window and extending/forecasting the signal. In the first class, judicious choice of analysis window whose support does not interfere with the boundary points can minimize the occurrence of aberrant patterns near the boundaries of the TF plane, for instance, \cite{Chui92wavelets,Depczynski99fast}. Due to the specific relationship between the chosen analysis windows and the TF analysis tool, these techniques do not make it possible to reduce the boundary effects of all TF representations. 
Another natural idea, the second class, consists of carrying out a preliminary step of extending, or forecasting, the signal beyond its boundaries. Due to its flexibility, various forecasting schemes have been proposed. For example, there exist simple extension schemes that do not take into account the dynamical behavior of the signal, such as zero-padding, periodic extension, symmetric extension~\cite{Kharitonenko02wavelet,Chen95symmetric}, or polynomial extrapolation~\cite{Williams97discrete}. 
There exist extension schemes based on physically relevant dynamical models, such as the Extended Dynamic Mode Decomposition~\cite{Williams15data} (EDMD) and the Gaussian process regression~\cite{Rasmussen06gaussian,Roberts13Gaussian} (GPR).
In speech processing, dynamic mode predictors have also been proposed~\cite{Vargas11speech} to forecast signals falling into the so-called \textit{source--filter} model. This gradient-based technique relies on the shadowing approach proposed in~\cite{Grebogi90shadowing}.
There also exist extension schemes based on stochastic models, such as the Trigonometric, Box-Cox transformation, ARMA errors, Trend and Seasonal components (TBATS) algorithm~\cite{DeLivera11forecasting}, the dynamic linear models \cite{west2006bayesian}. 
In the physically relevant dynamical models, the oscillation or trend, are usually modeled as the mean of a stochastic process, while in the stochastic models, the oscillation or trend are modeled by the covariance structure of the stochastic process. 
It is also possible to consider polynomial regression~\cite{fan1996local} or modeling the mean by splines~\cite{hall2005theory}, kernel functions~\cite{chang2010training}, and wavelets~\cite{marron1998exact}, or nonparametric regression~\cite{fan1996local}, to estimate the mean function before forecasting the signal. Neural network models, such as the long short-term memory model~\cite{vlachas2018data}, is another approach. 
The above list is far from exhaustive, and we refer readers with interest to \cite{hyndman2018forecasting} for a friendly monograph on the general forecasting topic. 
While the model-based extension scheme gives better-extended signals than those simple extension without dynamical models, they generally have a great computational cost. 

In this paper, we propose a fast extension algorithm based on a simple dynamical model that optimizes the trade-off between the extension quality and the computational cost, so that the real-time analysis can be achieved. The algorithm is composed of two steps.
\begin{enumerate}
\item \emph{Extend the signal by forecasting it.} The aim is to use a simple dynamic model to predict the values taken by the measured signal outside the measurement interval. Then, once this operation is done, we have access to an extended signal defined on a larger interval $I_\Delta$, where $\Delta$ denotes the size of the extension on both boundaries of $I$.
\item \emph{Run the local analysis tool on the extended signal.} Assuming that the support of the analysis window is smaller than $2\Delta$, the local analysis near the boundary of $I$ is now possible without lack of information thanks to knowledge brought by the extension. 
\end{enumerate}

Thus, assuming that the quality of the extension step is sufficient, the analysis results obtained that way will be less sensitive to the boundary effects than the result of the analysis tool applied directly to the nonextended signal. We claim and prove that forecasting oscillatory signals based on the simple dynamical model combined with the simple least square approach is sufficient for reducing the boundary effects for the TF analysis, or other kernel-based analysis. To our knowledge, such theoretical analysis does not exist in the literature. See the bottom of Fig.~\ref{fig:ex.intro}---in particular, the regions indicated by the dashed boxes---for a snapshot of the result. The main benefit of this simple approach is a numerically efficient solution with a theoretical guarantee for real-time analysis purposes. 

The paper is organized in the following way. In section~\ref{se:algo}, we provide an extension method based on a linear dynamic model. We derive the corresponding algorithm for boundary effects reduction. In Section~\ref{se:theoretical}, we show that the dynamic model we consider is sufficient to extend signals taking the form of sums of sine waves. An evaluation of the theoretical performance of our algorithm on this class of signals is given in Section~\ref{se:theoretical}. In Section~\ref{se:results}, we compare our extension method with more sophisticated methods such as EDMD, GPR, or TBATS. We show that our algorithm gives fast results of reasonable quality. Finally, we evaluate the performance of our boundary effects reduction algorithm on biomedical signals, such as respiratory signals, and compare it to the theoretical results. 

\section{Algorithm}
\label{se:algo}
\input{algorithm}

\section{Theoretical Performance}
\label{se:theoretical}
\input{theory}

\section{Numerical Results}
\label{se:results}
\input{results}

\section{Conclusion}
\label{se:conclusion}
In this paper, we propose an algorithm, named {\sf BoundEffRed}, for the real-time reduction of boundary effects in TF representations. This method is based on an extension of the signal obtained by a simple-minded and numerically efficient forecasting. We have shown theoretically that the chosen dynamic model is sufficient to extend signals formed by a sum of sine waves. Moreover, the low computational time allows us to switch to a real-time implementation of {\sf BoundEffRed}, unlike other existing forecasting methods. The numerical results also confirmed the robustness to noise of {\sf BoundEffRed}, as well as its ability to be applied to many TF representations. Additional applications to ECG and PPG are included in the Supplementary Materials.

Various improvements can be considered to make the algorithm more robust. In particular, we have noticed (see Fig.~\ref{fig:THO.failure}) that when the regular oscillations of the observed signal break, the forecasting step is no longer relevant, and only slows down the calculation of the TF representation. A preliminary step should then be added to the algorithm to detect signal activity and disable the forecasting step when possible. More fundamentally, one can also consider accelerating the computational time by optimizing the forecasting step to improve the real-time performance of {\sf BoundEffRed}. Indeed, each new forecast requires the inversion of the matrix $\bX\bX^T$ of size $M\times M$. However, the matrix $\bX$ used to forecast $\bx_{N+H}$ at the current iteration differs from the one used at the previous iteration to forecast $\bx_{N}$ by only $H$ columns. It then seems natural to take inspiration from the work of Strobach~\cite{Strobach97square}, generalized by Badeau \etal~\cite{Badeau04sliding}, which proposes a fast algorithm for the singular value decomposition of successive data matrices taking the same form as $\bX$. Their algorithm is limited to the case where $H=1$. Developing an extension of this algorithm to variations of $H>1$ columns would then allow us to efficiently update the matrix $\left(\bX\bX^T\right)^{-1}$. These avenues will be explored in our future work.

%\appendices
%\input{appendices}

%\section*{Acknowledgment}

\bibliographystyle{IEEEtran}
\bibliography{AM041320}

%\end{IEEEbiography}

\end{document}

\maketitle

\section{Proof of Theorem~1}
\label{ap:th.error}

\subsection{Preliminaries}

\subsubsection{Notations}
Consider $\bz\in\RR^N$ the deterministic signal defined in~(11) and denote by $\bz_k\in\RR^M$ the subsignal such that
\[
\bz_k[m] = \bz[N-K-M+k+m]\,,\quad \forall m\in\{0,\ldots,M-1\}\,,\ k\in\{0,\ldots,K\}\ .
\]
Define $\bZ\in\RR^{M \times K}$ and $\bZ'\in\RR^{M \times K}$, the matrices such that
\begin{align*}
\bZ &\defeq \begin{pmatrix}
\bz_0 & \cdots & \bz_{K-1}
\end{pmatrix}\,,\\ 
\bZ' &\defeq \begin{pmatrix}
\bz_1 & \cdots & \bz_{K}
\end{pmatrix}\,.
\end{align*}
Let $\bD\in\RR^{M\times M}$ be the matrix defined by $\bD[m,m']=\delta_{m+1,m'}$. Recall the model~(13). Based on the definition of matrices $\bX$ and $\bY$, we have:
\begin{align}
\label{eq:XX.T}
\dfrac1K\bX\bX^T &= \underbrace{\dfrac1K\bZ\bZ^T+\sigma^2\bI}_{\defeq \bS^{(0)}} + \bE^{(0)} \\
\label{eq:YX.T}
\dfrac1K\bY\bX^T &= \underbrace{\dfrac1K\bZ'\bZ^T +\sigma^2\bD}_{\defeq \bS^{(1)}}+ \bE^{(1)} \ ,
\end{align}
where $\bE^{(a)} \defeq \sigma\bE_1^{(a)} + \sigma^2\bE_2^{(a)}$,
\[
\bE_1^{(a)}[m,m'] = \dfrac1K\sum_{k=0}^{K-1} \bz[N_0+m+a+k]\bw[N_0+m'+k] + \bw[N_0+m+a+k]\bz[N_0+m'+k]\ ,
\]
and
\[
\bE_2^{(a)}[m,m'] =  \dfrac1K\sum_{k=0}^{K-1} \bw[N_0+m+a+k]\bw[N_0+m'+k] - \delta_{(m+a)m'}\ ,
\]
with $a\in\{0,1\}$.
We call $\bE^{(0)}$ and $\bE^{(1)}$ {\em error matrices} because:
\begin{align*}
\EE\{\bE^{(0)}\} &= \EE\{\bE_1^{(0)}\} = \EE\{\bE_2^{(0)}\} = \bzero \\
\EE\{\bE^{(1)}\} &= \EE\{\bE_1^{(1)}\} = \EE\{\bE_2^{(1)}\} = \bzero\ .
\end{align*}
Thus, the random matrix $\tilde\bA$, defined in equation~(8), is expressed in function of the above-defined matrices as:
\begin{equation*}
\tilde\bA = \left(\bS^{(1)}+\bE^{(1)}\right)\left(\bS^{(0)}+\bE^{(0)}\right)^{-1}\,.
\end{equation*}
Define $\bA_0$ the deterministic matrix such that
\begin{equation}
\bA_0 \defeq \bS^{(1)}{\bS^{(0)}}^\inv\,.
\label{eq:A0}
\end{equation}
We denote by $\balpha^{(\ell)}_0$ the last row of $\bA_0^\ell$. As a result, for $\ell\in \mathbb{N}^*$, the error vector $\bh^{(\ell)}$ defined by $\bh^{(\ell)}\defeq \balpha^{(\ell)} - \balpha^{(\ell)}_0$ satisfy the equation
\begin{align}
\nonumber
\bh^{(\ell)} &= \be_M^T\left(\tilde\bA^\ell-\bA_0^\ell\right) \\
&= \be_M^T\left(\left((\bS^{(1)}+\bE^{(1)})(\bS^{(0)}+\bE^{(0)})^\inv\right)^\ell- \bA_0^\ell\right)\ .
\label{eq:error.vec}
\end{align}

The randomness of $\bh^{(\ell)}$ completely comes from the error matrices. Besides, notice that the first $M-1$ rows in $\bE^{(1)}$ equal to the last $M-1$ rows of $\bE^{(0)}$. We gather all sources of randomness into a vector $\bg\in\RR^{M(M+1)}$, defined as
\begin{equation}
\bg = \vec\left(
\begin{bmatrix}
\bE^{(0)} \\
\be_M^T\bE^{(1)}
\end{bmatrix}
\right)\,,
\label{eq:vec}
\end{equation}
where "$\vec$" denotes the vectorization operator, that concatenates the columns of a given matrix on top of one another. Then, we can write $\bh^{(\ell)}$ as $\bh^{(\ell)}=f^{(\ell)}(\bg)$ where $f^{(\ell)}$ is a deterministic function such that:
\begin{align}
\nonumber
f^{(\ell)} : \RR^{M(M+1)} &\to \RR^{M} \\
 \bg &\mapsto \bh^{(\ell)}\ .
\label{eq:f.ell}
\end{align}

In the following paragraph, we provide some useful lemmas to prove Theorem 1.

\subsubsection{Lemmas}

\begin{lemma}[Expressions of $\bA_0$ and ${\bS^{(0)}}^{-1}$]
Let $\bS^{(0)}$ be the $M\times M$ matrix defined in~\eqref{eq:XX.T}. Let $\bA_0$ the $M\times M$ matrix defined in~\eqref{eq:A0}. Assume the deterministic signal $\bz$ takes the form~(11), and the observed noisy signal takes the form~(13). Then, the inverse of the matrix $\bS^{(0)}$ is given by
\begin{equation}
{\bS^{(0)}}^\inv[m,m']  = \dfrac1{\sigma^2}\delta_{m,m'}-\dfrac{2}{M\sigma^2}\sum_{j=1}^J\dfrac{1}{1+\frac{4\sigma^2}{M\Omega_j^2}}\cos\left(2\pi p_j \dfrac{m-m'}{M}\right)\,,
\label{eq:S0.inv}
\end{equation}
and the matrix $\bA_0$ is given by
\begin{align}
\bA_0[m,m']  &= \delta_{m+1,m'} + \dfrac2M\sum_{j=1}^J\dfrac{1}{1+\frac{4\sigma^2}{M\Omega_j^2}}\cos\left(2\pi p_j\frac{m'}{M}\right)\delta_{m+1,M}\,.
\label{eq:A0.sine}
\end{align}
Let $\|\cdot\|_{\max}$ denote the maximum norm of a matrix, \ie, $\|\bM\|_{\max} = \max_{n,n'} \left|\bM[n,n']\right|$. Then,
\begin{align}
\label{eq:S0.inv.bound}
\left\|{\bS^{(0)}}^\inv\right\|_{\max} &\leq \dfrac1{\sigma^2}\left(1+\frac{2J}{M}\right)\,, \\
\left\|\bA_0\right\|_{\max} &\leq \max\left(1,\frac{2J}{M}\right)\ .
\label{eq:A0.bound}
\end{align}
\end{lemma}

\begin{proof}
It follows from the signal model~(11) that the matrices $\bS^{(0)}$ and $\bS^{(1)}$ take the following form:
\begin{align}
\nonumber
\bS^{(a)}[m,m'] &= \sigma^2\delta_{(m+a)\,,m'}+\sum_{j,j'=1}^J\dfrac{\Omega_j\Omega_{j'}}{K}\sum_{k=0}^{K-1} \cos\left(2\pi \frac{f_j}{\fs}(N_0+m+a+k)+\varphi_j\right)\cos\left(2\pi \frac{f_{j'}}{\fs}(N_0+m'+k)+\varphi_{j'}\right) \\
\nonumber
&= \sigma^2\delta_{(m+a)\,,m'}+\sum_{j=1}^J\dfrac{\Omega_j^2}{2K}\sum_{k=0}^{K-1} \cos\left(2\pi \frac{f_j}{\fs}(m+a-m')\right) + \cos\left(2\pi \frac{f_j}{\fs}(2k+m+a+m'+2N_0)\right)\\
\nonumber
&= \sigma^2\delta_{(m+a)\,,m'}+\sum_{j=1}^J\Bigg( \dfrac{\Omega_j^2}2\cos\left(2\pi \frac{f_j}{\fs}(m+a-m')\right) + \dfrac{\Omega_j^2}{2K}\underbrace{\sum_{k=0}^{K-1}\cos\left(2\pi \frac{f_j}{\fs}(2k+m+a+m'+2N_0)\right)}_{=0\ \mathrm{because}\ \frac{f_j}\fs=\frac{p'_j}K } \Bigg)\\
&= \sigma^2\delta_{(m+a)\,,m'}+\sum_{j=1}^J\dfrac{\Omega_j^2}2\cos\left(2\pi \frac{f_j}{\fs}(m+a-m')\right)\ .
\label{eq:S1}
\end{align}
Thus, $\bS^{(0)}$ is a circulant matrix, and is therefore diagonalizable in the Fourier basis:
\[
\bS^{(0)} = \bU\bLambda^{(0)}\bU^*\ ,
\]
where $\bU[m,m']=\frac1{\sqrt{M}}e^{-2\ii\pi mm'/M}$ and $\bLambda^{(0)} = \mathrm{diag}(\lambda_0^{(0)},\dots,\lambda_{M-1}^{(0)})$ with
\begin{align*}
\lambda_m^{(0)} &= \sigma^2 + \sum_{j=1}^J\dfrac{\Omega_j^2}2\sum_{q=0}^{M-1} \cos\left(2\pi\frac{f_j}{\fs} q\right) e^{-2\ii\pi qm/M} \\
&= \sigma^2 + \dfrac{M}{4}\sum_{j=1}^J\Omega_j^2(\delta_{m,p_j} + \delta_{m,M-p_j})\ .
\end{align*}
Therefore,
\begin{align*}
{\bS^{(0)}}^\inv  &= \bU{\bLambda^{(0)}}^\inv\bU^*\,,
\end{align*}
which leads to
\begin{equation}
{\bS^{(0)}}^\inv[m,m']  = \dfrac1{\sigma^2}\delta_{m,m'}-\dfrac{2}{M\sigma^2}\sum_{j=1}^J\dfrac{1}{1+\frac{4\sigma^2}{M\Omega_j^2}}\cos\left(2\pi p_j \dfrac{m-m'}{M}\right)\,.
\label{eq:S0.inv.2}
\end{equation}
Directly, we have:
\begin{align*}
\left|{\bS^{(0)}}^\inv[m,m']\right|  &= \dfrac1{\sigma^2}\left|\delta_{m,m'}-\dfrac{2}{M}\sum_{j=1}^J\dfrac{1}{1+\frac{4\sigma^2}{M\Omega_j^2}}\cos\left(2\pi p_j \dfrac{m-m'}{M}\right)\right| \\
&\leq \dfrac1{\sigma^2}\left( 1+\dfrac{2}{M}\sum_{j=1}^J\dfrac{1}{1+\frac{4\sigma^2}{M\Omega_j^2}} \right) \leq \dfrac1{\sigma^2}\left( 1+\dfrac{2J}{M} \right)\ .
\end{align*}
Thus, 
\[
\left\|{\bS^{(0)}}^\inv\right\|_{\max}\leq\dfrac1{\sigma^2}\left( 1+\dfrac{2J}{M} \right)\ .
\]
Furthermore, combining equations~\eqref{eq:S1} and~\eqref{eq:S0.inv.2}, we have
\begin{align*}
\bA_0[m,m']  &= \sum_{q=0}^{M-1} {\bS^{(1)}}[m,q]{\bS^{(0)}}^\inv[q,m'] \\
&= \delta_{m+1,m'} + \dfrac2M\sum_{j=1}^J\dfrac{1}{1+\frac{4\sigma^2}{M\Omega_j^2}}\cos\left(2\pi p_j\frac{m'}{M}\right)\delta_{m+1,M}\,.
\end{align*}
Directly, we have:
\begin{align*}
\left|\bA_0[m,m']\right| &\leq \left\{
\begin{array}{ll}
1 & \mbox{if}\quad m<M-1\,, \\
\dfrac{2}{M}\sum_{j=1}^J\dfrac{1}{1+\frac{4\sigma^2}{M\Omega_j^2}} & \mbox{if}\quad m=M-1\ . \\
\end{array}
\right.
\end{align*}
Thus, 
\[
\left\|\bA_0\right\|_{\max}\leq \max\left(1,\frac{2J}{M}\right)\ .\qedhere
\]
\end{proof}

\begin{lemma}[Moments of $\bg$]
Let $\bg\in\RR^{M(M+1)}$ be the random vector defined in~\eqref{eq:vec}. Assume the deterministic signal $\bz$ takes the form~(11), and the observed noisy signal takes the form~(13). Then, as $K\to\infty$, the second-order moments of $\bg$ are bounded as follows:
\begin{equation}
\left|\bE\{\bg[r]\bg[r']\}\right| \leq \dfrac1K \left( C_\bz^2\sigma^2 + 2 \sigma^4\right) + o\left(\dfrac1K\right)\,,\quad\forall (r,r')\in\{0,\ldots,M(M+1)-1\}^2\ ,
\label{eq:gg.bound}
\end{equation}
where $C_\bz \defeq 2\left(\sum_{j=1}^J \Omega_j\right)$. Besides, higher-order moments of $\bg$ behave as $o\left( \dfrac1K \right)$.
\end{lemma}

\begin{proof}
In the following, for all $r\in\{0,\ldots,M(M+1)-1\}$, we denote $\bg[r] = \sigma\bg_1[r] + \sigma^2\bg_2[r]$, where 
\begin{align*}
\bg_1[r] &= \bE_1^{(a_r)}[m_r,m'_r] = \dfrac1K\sum_{k=0}^{K-1} \bz_k[m_r+a_r]\bw_k[m'_r] + \bw_k[m_r+a_r]\bz_k[m'_r]\,, \\
\bg_2[r] &= \bE_2^{(a_r)}[m_r,m'_r] = \dfrac1K\sum_{k=0}^{K-1} \bw_k[m_r+a_r]\bw_k[m'_r] - \delta_{m_r+a_r,\,m'_r} \,,
\end{align*}
and $m_r,m'_r, a_r$ are the corresponding coordinates of the matrices associated with $r$ through the vectorization operation~\eqref{eq:vec}. Thus, order-two moments of this random vector is split as follows:
\begin{equation}
\bE\{\bg[r]\bg[r']\} = \sigma^2\bE\{\bg_1[r]\bg_1[r']\} + \sigma^3\bE\{\bg_1[r]\bg_2[r']\} + \sigma^3\bE\{\bg_2[r]\bg_1[r']\} + \sigma^4\bE\{\bg_1[r]\bg_1[r']\}\ .
\label{eq:gg}
\end{equation}
By definition of the signal $\bz$ (see equation~(11)), we have $|\bz[n]|\leq\sum_{j=1}^J\Omega_j$, for all $n\in\NN$. Thus, by a direct bound, we have
\begin{align}
\nonumber
\left|\bE\{\bg_1[r]\bg_1[r']\}\right| & = \bigg|\dfrac1{K^2}\EE\bigg\{\sum_{k,k'=0}^{K-1} \bz_k[m_r+a_r]\bz_{k'}[m_{r'}+a_{r'}]\bw_k[m'_r]\bw_{k'}[m'_{r'}] + \bz_k[m'_r]\bz_{k'}[m_{r'}+a_{r'}]\bw_k[m_r+a_r]\bw_{k'}[m'_{r'}] \\
\nonumber
&+ \bz_k[m_r+a_r]\bz_{k'}[m'_{r'}]\bw_k[m'_r]\bw_{k'}[m_{r'}+a_{r'}] + \bz_k[m'_r]\bz_{k'}[m'_{r'}]\bw_k[m_r+a_r]\bw_{k'}[m_{r'}+a_{r'}]\bigg\} \bigg| \\
\nonumber
& \leq \dfrac1{K^2}\left(\sum_{j=1}^J \Omega_j\right)^2 \sum_{k,k'=0}^{K-1} \left|\EE\left\{\bw_k[m'_r]\bw_{k'}[m'_{r'}]\right\}\right| + \left|\EE\left\{\bw_k[m_r+a_r]\bw_{k'}[m'_{r'}]\right\}\right|  \\
&\hspace{110pt}  + \left|\EE\left\{\bw_k[m'_r]\bw_{k'}[m_{r'}+a_{r'}]\right\}\right| + \left|\EE\left\{\bw_k[m_r+a_r]\bw_{k'}[m_{r'}+a_{r'}]\right\}\right|
\label{eq:sum.g1g1}
\end{align}
Besides, since $\bw$ is a white noise,
\begin{align*}
\dfrac1{K^2}\sum_{k,k'=0}^{K-1} \left|\EE\left\{\bw_k[m'_r]\bw_{k'}[m'_{r'}]\right\}\right| & = \dfrac1{K^2}\sum_{k,k'=0}^{K-1} \delta_{k+m'_r,\,k'+m'_{r'}}  \\
&= \dfrac1{K^2}\left(K-\left|m'_r-m'_{r'}\right|\right) \ .
\end{align*}
Moreover, $0\leq\left|m'_r-m'_{r'}\right|\leq M-1$. Thus,
\begin{align*}
\dfrac1{K} - \dfrac{M-1}{K^2}\leq \dfrac1{K^2}\sum_{k,k'=0}^{K-1} \left|\EE\left\{\bw_k[m'_r]\bw_{k'}[m'_{r'}]\right\}\right| \leq \dfrac1{K}\ .
\end{align*}
Therefore,
\begin{align*}
\dfrac1{K^2}\sum_{k,k'=0}^{K-1} \left|\EE\left\{\bw_k[m'_r]\bw_{k'}[m'_{r'}]\right\}\right| = \dfrac1{K} + o\left(\dfrac1K\right)\ .
\end{align*}
Similar calculations lead to the same results for the other three terms making up the sum~\eqref{eq:sum.g1g1}.  Therefore, we have:
\begin{align}
\nonumber
\left|\bE\{\bg_1[r]\bg_1[r']\}\right| &\leq \left(\sum_{j=1}^J \Omega_j\right)^2 \left(\dfrac4{K} + o\left(\dfrac1{K}\right) \right) \\
&\leq \dfrac{C_\bz^2}{K} + o\left(\dfrac1{K}\right)\ .
\label{eq:g1g1}
\end{align}
Besides, since odd-order moments of a zero-mean multivariate Gaussian random vector are zero, we have:
\begin{align}
\nonumber
\bE\{\bg_1[r]\bg_2[r']\} & = \dfrac1{K^2}\sum_{k,k'=0}^{K-1} \bz_k[m_r+a_r]\EE\left\{\bw_k[m'_r]\bw_{k'}[m_{r'}+a_{r'}]\bw_{k'}[m'_{r'}]\right\} + \bz_k[m'_r]\EE\left\{\bw_k[m_r+a_r]\bw_{k'}[m_{r'}+a_{r'}]\bw_{k'}[m'_{r'}]\right\}\\
\nonumber
&\hspace{30pt} - \delta_{m_{r'}+a_{r'},m'_{r'}}\bE\{\bg_1[r]\} \\
& = 0\ .
\label{eq:g1g2}
\end{align}
Similarly,
\begin{equation}
\bE\{\bg_2[r]\bg_1[r']\} = 0\ .
\label{eq:g2g1}
\end{equation}
Besides, by a direct calculation, we have:
\begin{align}
\nonumber
\bE\{\bg_2[r]\bg_2[r']\} &= \dfrac1{K^2}\sum_{k,k'=0}^{K-1} \EE\left\{\bw_{k}[m_{r}+a_{r}]\bw_{k}[m'_{r}]\bw_{k'}[m_{r'}+a_{r'}]\bw_{k'}[m'_{r'}]\right\} - \delta_{m_{r}+a_{r},m'_{r}}\dfrac1{K}\sum_{k'=0}^{K-1} \EE\left\{\bw_{k'}[m_{r'}+a_{r'}]\bw_{k'}[m'_{r'}]\right\} \\
\nonumber
&\hspace{60pt} - \delta_{m_{r'}+a_{r'},m'_{r'}}\dfrac1{K}\sum_{k=0}^{K-1} \EE\left\{\bw_k[m'_r]\bw_{k'}[m'_{r'}]\right\} + \delta_{m_{r}+a_{r},m'_{r}}\delta_{m_{r'}+a_{r'},m'_{r'}} \\
&= \dfrac1{K^2}\sum_{k,k'=0}^{K-1} \EE\left\{\bw_{k}[m_{r}+a_{r}]\bw_{k}[m'_{r}]\bw_{k'}[m_{r'}+a_{r'}]\bw_{k'}[m'_{r'}]\right\} - \delta_{m_{r}+a_{r},m'_{r}}\delta_{m_{r'}+a_{r'},m'_{r'}}
\label{eq:sum.g2g2}
\end{align}
Moreover, using the results of the Isserlis' theorem~\cite{Isserlis16formula} to express fourth-order moments of a Gaussian random vector as a function of its second-order moments, we expand sum~\eqref{eq:sum.g2g2} as follows:
\begin{align*}
\bE\{\bg_2[r]\bg_2[r']\} &= \dfrac1{K^2} \sum_{k,k'=0}^{K-1} \EE\left\{\bw_{k}[m_{r}+a_{r}]\bw_{k}[m'_{r}]\right\}\EE\left\{\bw_{k'}[m_{r'}+a_{r'}]\bw_{k'}[m'_{r'}]\right\} - \delta_{m_{r}+a_{r},m'_{r}}\delta_{m_{r'}+a_{r'},m'_{r'}} \\
&\hspace{30pt}+ \dfrac1{K^2}\sum_{k,k'=0}^{K-1}\EE\left\{\bw_{k}[m_{r}+a_{r}]\bw_{k'}[m_{r'}+a_{r'}]\right\}\EE\left\{\bw_{k}[m'_{r}]\bw_{k'}[m'_{r'}]\right\} \\
&\hspace{30pt}+ \dfrac1{K^2}\sum_{k,k'=0}^{K-1} \EE\left\{\bw_{k}[m_{r}+a_{r}]\bw_{k'}[m'_{r'}]\right\}\EE\left\{\bw_{k}[m'_{r}]\bw_{k'}[m_{r'}+a_{r'}]\right\} \\
& = \dfrac1{K^2}\sum_{k,k'=0}^{K-1} \delta_{k+m_r+a_r,k'+m_{r'}+a_{r'}}\delta_{k+m'r,k'+m'_{r'}} + \delta_{k+m_r+a_r,k'+m'_{r'}}\delta_{k+m'_r,k'+m_{r'}+a_{r'}} \\
& = \dfrac1{K^2}\left(\delta_{m'_r-m'_{r'},m_r+a_r-m_{r'}-a_{r'}}(K-|m'_r-m'_{r'}|) +  \delta_{m_r+a_r-m'_{r'},m'_r-m_{r'}-a_{r'}}(K-|m_r+a_r-m'_{r'}|)\right) \\
& = \dfrac1{K}\left(\delta_{m'_r-m'_{r'},m_r+a_r-m_{r'}-a_{r'}} + \delta_{m'_r+m'_{r'},m_r+a_r+m_{r'}+a_{r'}}\right) + o\left( \dfrac1K\right)\ .
\end{align*}
Therefore,
\begin{equation}
\left| \bE\{\bg_2[r]\bg_2[r']\} \right| \leq \dfrac2{K} + o\left( \dfrac1K\right)\ .
\label{eq:g2g2}
\end{equation}
Thus, combining results~\eqref{eq:g1g1},~\eqref{eq:g1g2},~\eqref{eq:g2g1} and~\eqref{eq:g2g2} into expression~\eqref{eq:gg} gives the following bound:
\begin{equation*}
\left|\bE\{\bg[r]\bg[r']\}\right| \leq \dfrac1K \left( C_\bz^2\sigma^2 + 2 \sigma^4\right) + o\left(\dfrac1K\right)\ .
\end{equation*}

Concerning higher-order moments, let $T\geq 3$ denote the order of the moment defined by $\EE\{\prod_{\theta=1}^T\bg[r_\theta]\}$. Thus,
\begin{align}
\nonumber
\EE\left\{\prod_{\theta=1}^T\bg[r_\theta]\right\}& = \EE\left\{\prod_{\theta=1}^T \left(\sigma\bg_1[r_\theta] + \sigma^2\bg_2[r_\theta] \right)\right\}\\
\nonumber
&=\dfrac1{K^T}\EE\left\{\prod_{\theta=1}^T \left(\sum_{k=0}^{K-1}\rho_{\theta,k}\left(\bw[k+m'_{r_\theta}],\bw[k+m_{r_\theta}+a_{r_\theta}]\right) \right)\right\}\\ 
&=\dfrac1{K^T}\sum_{k_1=0}^{K-1}\cdots\sum_{k_T=0}^{K-1}\EE\left\{\prod_{\theta=1}^T \rho_{\theta,k_\theta}\left(\bw[k_\theta+m'_{r_\theta}],\bw[k_\theta+m_{r_\theta}+a_{r_\theta}]\right) \right\} \,,
\label{eq:multisum}
\end{align}
where
\[
\rho_{\theta,k}(u,v)=\sigma \bz_k[m_{r_\theta}+a_{r_\theta}]u + \sigma \bz_k[m'_{r_\theta}]v + \sigma^2uv - \delta_{m_{r_\theta}+a_{r_\theta},\,m'_{r_\theta}}\ .
\]
Thus,
\begin{equation*}
\left|\EE\left\{\prod_{\theta=1}^T\bg[r_\theta]\right\}\right| \leq \dfrac{\nu_K}{K^T} C_T\ ,
\end{equation*}
where
\[
C_T = \max_{(k_1,\ldots,k_T)} \left|\EE\left\{\prod_{\theta=1}^T \rho_{\theta,k_\theta}(\bw[k_\theta+m'_{r_\theta}],\bw[k_\theta+m_{r_\theta}+a_{r_\theta}]) \right\} \right| \,,
\]
and $\nu_K$ is the number of nonzero terms in the sum~\eqref{eq:multisum}. Note that $C_T$ is independent of $K$ (but depends on $\bz$ and $\sigma$). The behavior of the order $T$ moment in function of $K$ is therefore only determined by the ratio $\dfrac{\nu_K}{K^T}$.

Let us bound $\nu_K$. Fix $k_1$. For each of the other indexes of summation $k_\theta$ ($\theta\in\{2,\ldots,T\}$), there are four values that make $\rho_{\theta,k_\theta}(\bw[k_\theta+m'_{r_\theta}],\bw[k_\theta+m_{r_\theta}+a_{r_\theta}])$ not independent of $\rho_{1,k_1}(\bw[k_1+m'_{r_1}],\bw[k_1+m_{r_1}+a_{r_1}])$. Indeed, since $\bw$ is a white noise these quantities are independent except when $k_\theta$ is such that:
\begin{align*}
k_\theta &= k_1+m'_{r_1}-m'_{r_\theta} \\
k_\theta &= k_1+m_{r_1}+a_{r_1} - m'_{r_\theta}\\
k_\theta &= k_1+m'_{r_1} - m_{r_\theta}- a_{r_\theta} \\
k_\theta &= k_1+m_{r_1}+a_{r_1} - m_{r_\theta}- a_{r_\theta}\ .
\end{align*}
Consequently, for each value of $k_1$ there exist at least $(K-4)^{T-1}$ combinations of $(k_2,\ldots,k_T)$ where we have
\begin{align*}
\EE\left\{\prod_{\theta=1}^T \rho_{\theta,k_\theta}(\bw[k_\theta+m'_{r_\theta}],\bw[k_\theta+m_{r_\theta}+a_{r_\theta}]) \right\} &= \EE\left\{ \rho_{1,k_1}(\bw[k_1+m'_{r_1}],\bw[k_1+m_{r_1}+a_{r_1}]) \right\} \\
&\hspace{20pt}\times\EE\left\{\prod_{\theta=2}^T \rho_{\theta,k_\theta}(\bw[k_\theta+m'_{r_\theta}],\bw[k_\theta+m_{r_\theta}+a_{r_\theta}]) \right\}
& = 0\,,
\end{align*}
because $\EE\left\{ \rho_{1,k_1}(\bw[k_1+m'_{r_1}],\bw[k_1+m_{r_1}+a_{r_1}]) \right\}=0$. Therefore, at least $K(K-4)^{T-1}$ of the sum~\eqref{eq:multisum} are zero. Because $T\geq 3$, we develop similar arguments on $k_2$ and $k_3$ to determine other cases where this correlation term vanishes. We subtract these cases to $K^T$, the total number of combinations of $(k_1,\ldots,k_T)$ to obtain the following maximum bound on the number of nonzero terms in the sum~\eqref{eq:multisum}:
\[
\nu_K\leq K^T - 3K(K-4)^{T-1} + 3K(K-4)(K-8)^{T-2}- K(K-4)(K-8)(K-12)^{T-3}\ .
\]
Thus,
\begin{align*}
\left|\EE\left\{\prod_{\theta=1}^T\bg[r_\theta]\right\}\right| &\leq C_T \dfrac{K^T - 3K(K-4)^{T-1} + 3K(K-4)(K-8)^{T-2}- K(K-4)(K-8)(K-12)^{T-3}}{K^T} \\
&\leq C_T \left( 1 - 3\left(1-\frac4K\right)^{T-1} + 3\left(1-\frac4K\right)\left(1-\frac8K\right)^{T-2}- \left(1-\frac4K\right)\left(1-\frac8K\right)\left(1-\frac{12}{K}\right)^{T-3} \right) \\
&\leq C_T \left( 1 - 3 +\frac{12(T-1)}{K} + 3-\frac{12}K-\frac{24(T-2)}{K} -1 + \frac{4}K + \frac{8}K +\frac{12(T-3)}{K} + o\left(\dfrac1K\right)\right) \\
&\leq C_T\ o\left(\dfrac1K\right)\ . \\
\end{align*}
Therefore,
\begin{equation*}
\left|\EE\left\{\prod_{\theta=1}^T\bg[r_\theta]\right\}\right| = o\left(\dfrac1K\right)\,,\quad\forall T\geq 3\ .\qedhere
\end{equation*}
\end{proof}

\begin{lemma}[Bounds on the derivatives of $f^{(\ell)}$ at the origin]
Let $f^{(\ell)} : \RR^{M(M+1)} \to \RR^{M}$ denote the multivariate function defined in~\eqref{eq:f.ell}. Assume the deterministic signal $\bz$ takes the form~(11), and the observed noisy signal takes the form~(13). Then, the first-order derivatives of $f^{(\ell)}$ at the origin are bounded as follows:  
\begin{equation}
\left|\left.\dfrac{\partial f_m^{(\ell)}}{\partial \bg[r]}\right\vert_{\bg=\bzero}\right| \leq \dfrac{d_{1,\bz,M,\ell}}{\sigma^2}\,,\quad\forall m\in\{0,\ldots,M-1\}\,,\ r\in\{0,\ldots,M(M+1)-1\}\ ,
\label{eq:df.bound}
\end{equation}
where
\[
d_{1,\bz,M,\ell}\defeq \left(2 + (\ell-2)M\right)M^{\ell-1}\left(\max\left(1,\frac{2J}{M}\right)\right)^{\ell-1}\left(1+\max\left(1,\frac{2J}{M}\right)\right)\left(1+\frac{2J}{M}\right)\ .
\]
Besides, the second-order derivatives of $f^{(\ell)}$ at the origin are bounded as follows:  
\begin{equation}
\left|\left.\dfrac{\partial^2 f_m^{(\ell)}}{\partial \bg[r]\partial \bg[r']}\right\vert_{\bg=\bzero}\right| \leq \dfrac{d_{2,\bz,M,\ell}}{\sigma^4}\,,\quad\forall m\in\{0,\ldots,M-1\}\,,\ (r,r')\in\{0,\ldots,M(M+1)-1\}^2\ ,
\label{eq:d2f.bound}
\end{equation}
where 
\begin{align*}
d_{2,\bz,M,\ell} &\defeq \left(1+\frac{2J}{M}\right)^2 \left(\max\left(1,\frac{2J}{M}\right)\right)^{\ell-2}\left(1+\max\left(1,\frac{2J}{M}\right)\right)\left(d_{2,M,\ell} + (d_{2,M,\ell}+2d'_{2,M,\ell})\max\left(1,\frac{2J}{M}\right)\right)\,,
\end{align*}
and $d_{2,M,\ell}$ and $d'_{2,M,\ell}$ are only depending on $M$ and $\ell$.
\end{lemma}

\begin{proof}
Concerning the first-order derivative, from~\eqref{eq:error.vec}, we have:
\begin{align*}
\dfrac{\partial f^{(\ell)}}{\partial \bg[r]} &= \be_M^T\dfrac{\partial \tilde\bA^{\ell}}{\partial \bg[r]} \\
&= \sum_{\lambda=0}^{\ell-1} \be_M^T\tilde\bA^\lambda\dfrac{\partial \tilde\bA}{\partial \bg[r]}\tilde\bA^{\ell-1-\lambda}\ .
\end{align*}
Thus,
\begin{equation}
\left.\dfrac{\partial f^{(\ell)}}{\partial \bg[r]}\right|_{\bg=\bzero} = \sum_{\lambda=0}^{\ell-1} \be_M^T\bA_0^\lambda\left.\dfrac{\partial \tilde\bA}{\partial \bg[r]}\right|_{\bg=\bzero}\bA_0^{\ell-1-\lambda}\ .
\label{eq:df}
\end{equation}
Furthermore,
\begin{align*}
\dfrac{\partial \tilde\bA}{\partial \bg[r]} &= \dfrac{\partial \bE^{(1)}}{\partial \bg[r]}\left( \bS^{(0)}+\bE^{(0)}\right)^{-1} +  \left(\bS^{(1)}+\bE^{(1)}\right)\dfrac{\partial \left( \bS^{(0)}+\bE^{(0)}\right)^{-1}}{\partial \bg[r]} \\
&= \left\{
\begin{array}{ll}
\bJ_{m_r,m'_r}\left( \bS^{(0)}+\bE^{(0)}\right)^{-1} & \mbox{if}\quad a_r=1, \\
\left( (1-\delta_{m_r,0})\bJ_{m_r-1,m'_r} +\left(\bS^{(1)}+\bE^{(1)}\right)\left( \bS^{(0)}+\bE^{(0)}\right)^{-1}\bJ_{m_r,m'_r} \right) \left( \bS^{(0)}+\bE^{(0)}\right)^{-1} & \mbox{else,}
\end{array}
\right.
\end{align*}
where $\bJ_{m_r,m'_r}\in\RR^{M\times M}$ is the matrix such that $\bJ_{m_r,m'_r}[m,m'] \defeq \delta_{m,m_r}\delta_{m',m'_r}$.
Thus,
\begin{align*}
\left.\dfrac{\partial \tilde\bA}{\partial \bg[r]}\right|_{\bg=\bzero} &= \left\{
\begin{array}{ll}
\bJ_{m_r,m'_r}{\bS^{(0)}}^{-1} & \mbox{if}\quad a_r=1, \\
\left((1-\delta_{m_r,0})\bJ_{m_r-1,m'_r} +\bA_0\bJ_{m_r,m'_r}\right){\bS^{(0)}}^{-1} & \mbox{else.}
\end{array}
\right.
\end{align*}
Then,
\begin{equation*}
\left\| \left.\dfrac{\partial \tilde\bA}{\partial \bg[r]}\right|_{\bg=\bzero}\right\|_{\max} \leq \left(1+\|\bA_0\|_{\max}\right)\left\|{\bS^{(0)}}^{-1}\right\|_{\max}\ .
\end{equation*}
Given bounds~\eqref{eq:A0.bound} and~\eqref{eq:S0.inv.bound}, we have that:
\begin{equation}
\left\| \left.\dfrac{\partial \tilde\bA}{\partial \bg[r]}\right|_{\bg=\bzero}\right\|_{\max} \leq \left(1+\max\left(1,\frac{2J}{M}\right)\right)\left(1+\frac{2J}{M}\right)\dfrac{1}{\sigma^2}\ .
\label{eq:dA}
\end{equation}
Besides given expression~\eqref{eq:df}, for all $r\in\{0,\ldots,M(M+1)-1\}$, we have:
\begin{align*}
\left|\left.\dfrac{\partial f_m^{(\ell)}}{\partial \bg[r]}\right\vert_{\bg=\bzero}\right| &\leq 2M\|\bA_0^{\ell-1}\|_{\max} \left\| \left.\dfrac{\partial \tilde\bA}{\partial \bg[r]}\right|_{\bg=\bzero}\right\|_{\max} + M^2\sum_{\lambda=1}^{\ell-2} \|\bA_0^\lambda\|_{\max} \left\| \left.\dfrac{\partial \tilde\bA}{\partial \bg[r]}\right|_{\bg=\bzero}\right\|_{\max}  \|\bA_0^{\ell-1-\lambda}\|_{\max} \\
&\leq \left(2 + (\ell-2)M\right)M^{\ell-1} \|\bA_0\|_{\max}^{\ell-1}\left\|\left.\dfrac{\partial\tilde\bA}{\partial \bg[r]}\right|_{\bg=\bzero}\right\|_{\max}\ .
\end{align*}
Therefore, given bounds~\eqref{eq:A0.bound} and~\eqref{eq:dA}, we have:
\begin{equation*}
\left|\left.\dfrac{\partial f_m^{(\ell)}}{\partial \bg[r]}\right\vert_{\bg=\bzero}\right| \leq \dfrac{d_{1,\bz,M,\ell}}{\sigma^2}\ .
\end{equation*} 

Concerning the second-order derivative, we have:
\begin{align*}
\dfrac{\partial^2 f^{(\ell)}}{\partial \bg[r]\partial \bg[r']} &= \sum_{\lambda=0}^{\ell-1} \be_M^T\dfrac{\partial\tilde\bA^\lambda}{\partial \bg[r']}\dfrac{\partial \tilde\bA}{\partial \bg[r]}\tilde\bA^{\ell-1-\lambda} + \be_M^T\tilde\bA^\lambda\dfrac{\partial^2 \tilde\bA}{\partial \bg[r]\partial \bg[r']}\tilde\bA^{\ell-1-\lambda} + \be_M^T\tilde\bA^\lambda\dfrac{\partial \tilde\bA}{\partial \bg[r]}\dfrac{\partial\tilde\bA^{\ell-1-\lambda}}{\partial \bg[r']} \\
&= \sum_{\lambda=1}^{\ell-1}\sum_{p=0}^{\lambda-1} \be_M^T\tilde\bA^p\dfrac{\partial\tilde\bA}{\partial \bg[r']}\tilde\bA^{\lambda-1-p}\dfrac{\partial \tilde\bA}{\partial \bg[r]}\tilde\bA^{\ell-1-\lambda} + \sum_{\lambda=0}^{\ell-1}\be_M^T\tilde\bA^l\dfrac{\partial^2 \tilde\bA}{\partial \bg[r]\partial \bg[r']}\tilde\bA^{\ell-1-\lambda} \\ 
&+ \sum_{\lambda=0}^{\ell-2}\sum_{p=0}^{\ell-\lambda-2}\be_M^T\tilde\bA^\lambda\dfrac{\partial \tilde\bA}{\partial \bg[r]}\tilde\bA^p\dfrac{\partial\tilde\bA}{\partial \bg[r']}\tilde\bA^{\ell-\lambda-2-p}\ . \\
\end{align*}
Thus,
\begin{align}
\nonumber
\left.\dfrac{\partial^2 f^{(\ell)}}{\partial \bg[r]\partial \bg[r']}\right\vert_{\bg=\bzero} &= \sum_{\lambda=1}^{\ell-1}\sum_{p=0}^{\lambda-1} \be_M^T\bA_0^{p}\left.\dfrac{\partial\tilde\bA}{\partial \bg[r']}\right\vert_{\bg=\bzero}\bA_0^{\lambda-1-p}\left.\dfrac{\partial \tilde\bA}{\partial \bg[r]}\right\vert_{\bg=\bzero}\bA_0^{\ell-1-\lambda} \\
&+ \sum_{\lambda=0}^{\ell-1}\be_M^T\bA_0^{\lambda}\left.\dfrac{\partial^2 \tilde\bA}{\partial \bg[r]\partial \bg[r']}\right\vert_{\bg=\bzero}\bA_0^{\ell-1-\lambda} + \sum_{\lambda=0}^{\ell-2}\sum_{p=0}^{\ell-2-\lambda}\be_M^T\bA_0^{\lambda}\left.\dfrac{\partial \tilde\bA}{\partial \bg[r]}\right\vert_{\bg=\bzero}\bA_0^p\left.\dfrac{\partial\tilde\bA}{\partial \bg[r']}\right\vert_{\bg=\bzero}\bA_0^{\ell-\lambda-2-p}\ .
\label{eq:d2f}
\end{align}
Besides, the second-order derivative of the matrix $\tilde\bA$ is given by
\begin{align*}
\dfrac{\partial^2 \tilde\bA}{\partial \bg[r]\partial \bg[r']} &= \left\{
\begin{array}{lc}
0 &\mbox{if}\quad a_r=1\quad\mbox{and}\quad {a_r'}=1, \\
\bJ_{m_{r},m'_{r}}\left( \bS^{(0)}+\bE^{(0)}\right)^{-1} \bJ_{m_{r'},m'_{r'}} \left( \bS^{(0)}+\bE^{(0)}\right)^{-1} & \mbox{if}\quad a_r=1\quad\mbox{and}\quad {a_r'}=0, \\
\bJ_{m'_r,m'_{r'}}\left( \bS^{(0)}+\bE^{(0)}\right)^{-1}\bJ_{m_r,m'_r}\left( \bS^{(0)}+\bE^{(0)}\right)^{-1} & \mbox{if}\quad a_r=0\quad\mbox{and}\quad {a_r'}=1, \\
\left( (1-\delta_{m_r,0})\bJ_{m_r-1,m'_r} +\left(\bS^{(1)}+\bE^{(1)}\right)\left( \bS^{(0)}+\bE^{(0)}\right)^{-1}\bJ_{m_r,m'_r} \right) & \\
\hspace{20pt}\times \left( \bS^{(0)}+\bE^{(0)}\right)^{-1}\bJ_{m_{r'},m'_{r'}}\left( \bS^{(0)}+\bE^{(0)}\right)^{-1} & \\
+ \left( (1-\delta_{m_{r'},0})\bJ_{m_{r'}-1,m'_{r'}} + \left(\bS^{(1)}+\bE^{(1)}\right)\left( \bS^{(0)}+\bE^{(0)}\right)^{-1}\bJ_{m_{r'},m'_{r'}} \right) & \\
\hspace{20pt}\times \left( \bS^{(0)}+\bE^{(0)}\right)^{-1}\bJ_{m_{r},m'_{r}}\left( \bS^{(0)}+\bE^{(0)}\right)^{-1}  & \mbox{else.}
\end{array} 
\right.
\end{align*}
Thus, the second-order derivative of the matrix $\tilde\bA$ at the origin is such that
\begin{align*}
\left.\dfrac{\partial^2 \tilde\bA}{\partial \bg[r]\partial \bg[r']}\right|_{\bg=\bzero} &= \left\{
\begin{array}{ll}
0 &\mbox{if}\quad a_r=1\quad\mbox{and}\quad a_{r'}=1, \\
\bJ_{m_r,m'_r}{\bS^{(0)}}^{-1} \bJ_{m_{r'},m'_{r'}} { \bS^{(0)}}^{-1} & \mbox{if}\quad a_r=1\quad\mbox{and}\quad a_{r'}=0, \\
\bJ_{m_{r'},m'_{r'}}{\bS^{(0)}}^{-1} \bJ_{m_{r},m'_{r}} { \bS^{(0)}}^{-1} & \mbox{if}\quad a_r=0\quad\mbox{and}\quad a_{r'}=1, \\
\left( (1-\delta_{m_r,0})\bJ_{m_r-1,m'_r} +\bA_0\bJ_{m_r,m'_r} \right){\bS^{(0)}}^{-1}\bJ_{m_{r'},m'_{r'}}{ \bS^{(0)}}^{-1} & \\
+ \left( (1-\delta_{m_{r'},0})\bJ_{m_{r'}-1,m'_{r'}} +\bA_0\bJ_{m_{r'},m'_{r'}} \right){\bS^{(0)}}^{-1}\bJ_{m_{r},m'_{r}}{ \bS^{(0)}}^{-1} & \mbox{else.}
\end{array} 
\right.
\end{align*}
Then,
\begin{align}
\nonumber
\left\| \left.\dfrac{\partial^2 \tilde\bA}{\partial \bg[r]\partial \bg[r']}\right|_{\bg=0}\right\|_{\max} &\leq \left\{
\begin{array}{ll}
\left\|{\bS^{(0)}}^{-1}\right\|_{\max}^2 &\mbox{if}\quad a_r=1\quad\mbox{or}\quad a_{r'}=1\,, \\
2\left(1+\|\bA_0\|_{\max}\right)\left\|{\bS^{(0)}}^{-1}\right\|_{\max}^2 &\mbox{else} \\
\end{array}
\right.\\
&\leq 2\left(1+\max\left(1,\frac{2J}{M}\right)\right)\left(1+\frac{2J}{M}\right)^2\dfrac1{\sigma^4}\ .
\label{eq:d2A}
\end{align}
Returning to equation~\eqref{eq:d2f}, for all $r,r'\in\{0,\ldots,M(M+1)-1\}$ and $m\in\{0,\ldots, M-1\}$, we have:
\begin{align*}
\left|\left.\dfrac{\partial^2 f_m^{(\ell)}}{\partial \bg[r]\partial \bg[r']}\right\vert_{\bg=\bzero}\right| &\leq d_{2,M,\ell} \|\bA_0\|_{\max}^{\ell-2}\left\|\dfrac{\partial\tilde\bA}{\partial \bg[r]}\right\|_{\max}\left\|\dfrac{\partial\tilde\bA}{\partial \bg[r']}\right\|_{\max} + d'_{2,M,\ell}\|\bA_0\|_{\max}^{\ell-1}\left\|\dfrac{\partial^2\tilde\bA}{\partial \bg[r]\bg[r']}\right\|_{\max} \ ,
\end{align*}
where $d_{2,M,\ell}$ and $d'_{2,M,\ell}$ are only depending on $M$ and $\ell$.
Besides, given results~\eqref{eq:A0}, ~\eqref{eq:dA} and~\eqref{eq:d2A}, we have:
\begin{equation*}
\left|\left.\dfrac{\partial^2 f_m^{(\ell)}}{\partial \bg[r]\partial \bg[r']}\right\vert_{\bg=\bzero}\right| \leq \dfrac{d_{2,\bz,M,\ell}}{\sigma^4}\ .\qedhere
\end{equation*}
\end{proof}

\subsection{Expression of the Bias $\bmu$.}

By definition of the measurement noise, $\bmu[n]=0$ when $n\in I$. Outside the measurement interval $I$, denote by $\ell$ the index such that $n=N-1+\ell$. Then, given that $\bh^{(\ell)} = \balpha^{(\ell)} - \balpha_0^{(\ell)}$, we deduce from expression~(16) that
\begin{align}
\nonumber
\bmu[n] &=  \EE\{\balpha^{(\ell)}\}\bz_{K} + \sigma\EE\{\balpha^{(\ell)}\bw_{K}\} - \bz[n]\\
\nonumber
& = \balpha_0^{(\ell)}\bz_{K} + \EE\{\bh^{(\ell)}\}\bz_{K} + \sigma\EE\{\bh^{(\ell)}\bw_{K}\} - \bz[N-1+\ell] \\
&\defeq \bepsilon_1[\ell] + \bepsilon_2[\ell] + \bepsilon_3[\ell]\ ,
\label{eq:bias.dec}
\end{align}

where
\begin{align}
\label{eq:eps1.def}
\bepsilon_1[\ell]&=  \balpha_0^{(\ell)}\bz_{K}  - \bz[N-1+\ell]\,, \\
\label{eq:eps2.def}
\bepsilon_2[\ell] &=  \EE\{\bh^{(\ell)}\}\bz_{K}\,,  \\
\label{eq:eps3.def}
\bepsilon_3[\ell] &=  \sigma\EE\{\bh^{(\ell)}\bw_{K}\}  \ .
\end{align}

Let us first determine an upper bound on $\left|\bepsilon_1[n]\right|$.  Since $\balpha_0^{(1)}$ is the last row of $\bA_0$, we deduce from the expression~\eqref{eq:A0.sine} of $\bA_0$ that
\begin{equation}
\balpha_0^{(1)}[m]  =\dfrac2M \sum_{j=1}^J\dfrac{1}{1+\frac{4\sigma^2}{M\Omega_j^2}}\cos\left(2\pi p_j\frac{m}{M}\right) \ . 
\label{eq:alpha01}
\end{equation}
Besides, from equation~\eqref{eq:A0.sine}, we also have
\begin{align*}
\bA_0\bz_{K} &= 
\begin{pmatrix}
\bz[N-M+1] \\
\vdots \\
\bz[N-1] \\
\balpha_0^{(1)}\bz_K
\end{pmatrix}\,.
\end{align*}
The upward-shift property is thus successively inducted when $\ell$ increases; that is,
\begin{align*}
\bA_0^{\ell}\bz_{K} &= 
\begin{pmatrix}
\bz[N-M+\ell] \\
\vdots \\
\bz[N-1] \\
\balpha_0^{(1)}\bz_K \\
\vdots \\
\balpha_0^{(\ell)}\bz_K
\end{pmatrix}
\, .
\end{align*}
Then, $\balpha_0^{(\ell)}$, the last row of $\bA_0^\ell$ follows the following recurrence relation:
\begin{align*}
\balpha_0^{(\ell)}\bz_{K} &= \balpha_0^{(1)}\tilde\bA_0^{\ell-1}\bz_{K} \\
&=\sum_{m=0}^{M-\ell}\balpha_0^{(1)}[m]\bz[N-M+\ell+m-1]+\sum_{m=M-\ell+1}^{M-1}\balpha_0^{(1)}[m]\balpha_0^{(m-M+\ell)}\bz_K\,.
\end{align*}
Hence,
\begin{align*}
\bepsilon_1[\ell] &= \balpha_0^{(\ell)}\bz_{K} - \bz[N-1+\ell] \\
&=\sum_{m=0}^{M}\balpha_0^{(1)}[m]\bz[N-M+\ell+m-1]  - \bz[N-1+\ell] +\sum_{m=M-\ell+1}^{M-1}\balpha_0^{(1)}[m]\left(\balpha_0^{(m-M+\ell)}\bz_K - \bz[N-M+\ell+m-1]\right) \\
&=\sum_{m=0}^{M}\balpha_0^{(1)}[m]\bz[N-M+\ell+m-1]  - \bz[N-1+\ell] +\sum_{m=M-\ell+1}^{M-1}\balpha_0^{(1)}[m]\bepsilon_1[m-M+\ell]\,.
\end{align*}
Besides, equation~\eqref{eq:alpha01} gives
\begin{align*}
\sum_{m=0}^{M}\balpha_0^{(1)}[m]\bz[N-M+\ell+m-1] &=\sum_{j,j'=1}^J\Omega_{j'}\dfrac{2}{M}\dfrac1{1+\frac{4\sigma^2}{M\Omega_j^2}}\underbrace{\sum_{m=0}^{M-1}\cos\left(2\pi p_j\frac{m}{M}\right)\cos\left(2\pi p_{j'}\dfrac{N+m}{M}+\varphi_{j'}\right)}_{=\delta_{j,j'}\frac{M}2\cos\left(2\pi p_j\frac{N}M +\varphi_{j}\right)} \\
&= \sum_{j=1}^J\Omega_j\dfrac1{1+\frac{4\sigma^2}{M\Omega_j^2}}\cos\left(2\pi p_j\dfrac{N}{M}+\varphi_j\right)\ .
\end{align*}
Thus:
\begin{align}
\nonumber
\left|\bepsilon_1[\ell]\right| &= \left|\sum_{j=1}^J\Omega_j\left(\dfrac1{1+\frac{4\sigma^2}{M\Omega_j^2}}-1\right)\cos\left(2\pi p_j\dfrac{N}{M}+\varphi_j\right) +\sum_{m=M-\ell+1}^{M-1}\balpha_0^{(1)}[m]\bepsilon_1[m-M+\ell]\right| \\
\nonumber
&\leq \sum_{j=1}^J\Omega_j\left|\dfrac1{1+\frac{4\sigma^2}{M\Omega_j^2}}-1\right| +\dfrac{2J}{M}\sum_{\lambda=1}^{\ell-1}\left|\bepsilon_1[\lambda]\right| \\
&\leq \dfrac{4\sigma^2}{M}\sum_{j=1}^J\dfrac1{\Omega_j} +\dfrac{2J}{M}\sum_{\lambda=1}^{\ell-1}\left|\bepsilon_1[\lambda]\right|\ .
\label{eq:seq}
\end{align}
Then, by induction from the inequality~\eqref{eq:seq}, we have
\begin{equation}
\left|\bepsilon_1[\ell]\right| \leq \dfrac{4\sigma^2}{M}\left(1+\frac{2J}{M}\right)^{\ell-1}\sum_{j=1}^J\dfrac1{\Omega_j} \defeq c^{(1)}\,\sigma^2\ ,
\label{eq:eps1}
\end{equation}
where $c^{(1)}=\dfrac{4}{M}\left(1+\frac{2J}{M}\right)^{\ell-1}\sum_{j=1}^J\dfrac1{\Omega_j}$. Note that $c^{(1)}$ is not depending on $\sigma$ or $K$.

Let us now determine an upper bound on $\left|\bepsilon_2[\ell]\right|$. Since $\EE\{\bg[r]\}=0$ and moments of order $3$ and higher behave as $o\left(\dfrac1K\right)$, a second-order Taylor expansion of $\bh^{(\ell)}$ gives
\begin{equation}
\EE\{\bh^{(\ell)}[m]\} = \dfrac12\sum_{r,r'=0}^{M(M+1)-1}\left.\dfrac{\partial^2 f_m^{(\ell)}}{\partial \bg[r]\partial \bg[r']}\right\vert_{\bg=\bzero} \EE\{\bg[r]\bg[r']\} + o\left(\dfrac1K\right)\ .
\label{eq:Eh}
\end{equation}
Thus, given the bounds~\eqref{eq:gg.bound} on $\left|\EE\{\bg[r]\bg[r']\}\right|$ and~\eqref{eq:d2f.bound} on the second derivative of $f_m^{(\ell)}$, we have:
\begin{align}
\nonumber
\left|\bepsilon_2[\ell]\right| &\leq C_\bz\dfrac{M^3(M+1)^2}{4}\left\|\left.\dfrac{\partial^2 f^{(\ell)}}{\partial \bg[r]\partial \bg[r']}\right\vert_{\bg=\bzero}\right\|_{\max} \dfrac1K \left( C_\bz^2\sigma^2 + 2 \sigma^4\right) + o\left(\dfrac1K\right)  \\
&\leq \dfrac1K \left(c^{(2)}_1 + \dfrac{c^{(2)}_2}{\sigma^2} \right) + o\left(\dfrac1K\right)\ ,
\label{eq:eps2}
\end{align}
where
\begin{align*}
c^{(2)}_1 &\defeq\dfrac{M^3(M+1)^2}{2}C_\bz d_{2,\bz,M,\ell}\ ,\\
c^{(2)}_2 &\defeq\dfrac{M^3(M+1)^2}{4}C_\bz^3 d_{2,\bz,M,\ell} \ .
\end{align*}

Let us now determine an upper bound on $\left|\bepsilon_3[\ell]\right|$. 
A second-order Taylor expansion of $\bh^{(\ell)}$ gives
\begin{equation}
\EE\{\bh^{(\ell)}\bw_K\}  = \sum_{m=0}^{M-1} \sum_{r=0}^{M(M+1)-1}\left.\dfrac{\partial f_m^{(\ell)}}{\partial \bg[r]}\right\vert_{\bg=\bzero} \EE\{\bg[r]\bw_K[m]\} + o\left(\dfrac1K\right)
\label{eq:Ehw}
\end{equation}
Indeed, the third-order moments $\EE\{\bg[r]\bg[r']\bw_K[m]\}$, behave as $o\left(\dfrac1K\right)$. Besides,
\begin{equation}
\EE\{\bg[r]\bw_K[m]\}= \sigma\EE\{\bg_1[r]\bw_K[m]\} + \sigma^2\EE\{\bg_2[r]\bw_K[m]\}\ .
\label{eq:gw}
\end{equation}
Then,
\begin{align}
\nonumber
\left|\bE\{\bg_1[r]\bw_K[m]\}\right| & = \dfrac1{K}\left|\sum_{k=0}^{K-1} \bz_k[m_r+a_r]\EE\left\{\bw_k[m'_r]\bw_{K}[m]\right\} + \bz_k[m'_r]\EE\left\{\bw_k[m_r+a_r]\bw_{K}[m]\right\}\right| \\
& \leq \dfrac2{K}\left(\sum_{j=1}^J \Omega_j\right) = \dfrac{C_\bz}{K}\ .
\label{eq:g1w}
\end{align}
\begin{align}
\nonumber
\bE\{\bg_2[r]\bw_K[m]\} & = \dfrac1{K}\sum_{k,=0}^{K-1} \EE\left\{\bw_{k}[m_{r}+a_{r}]\bw_{k}[m'_{r}]\bw_K[m]\right\} -\delta_{m_r+a_r,m'_r}\EE\left\{\bw_K[m]\right\}\\
& = 0\ .
\label{eq:g2w}
\end{align}
Thus, combining results~\eqref{eq:g1w} and~\eqref{eq:g2w} into expression~\eqref{eq:gw} gives the following bound:
\begin{equation}
\left|\bE\{\bg[r]\bw_K[m]\}\right| \leq C_\bz\dfrac{\sigma}K\ .
\label{eq:gw.bound}
\end{equation}
Thus, given the bound~\eqref{eq:df.bound} on the first derivative of $f_m^{(\ell)}$, and from~\eqref{eq:Ehw} we have:
\begin{equation}
\left|\bepsilon_3[\ell]\right| \leq \sigma M^2(M+1) \dfrac{d_{1,\bz,M,\ell}}{\sigma^2} C_\bz \dfrac{\sigma}{K} \defeq \dfrac{c^{(3)}}{{K}}\ ,
\label{eq:eps3}
\end{equation}
where $ c^{(3)} \defeq M^2(M+1)d_{1,\bz,M,\ell}C_\bz$.

Thus, combining bounds~\eqref{eq:eps1} on $\bepsilon_1$, ~\eqref{eq:eps2} on $\bepsilon_2$ and ~\eqref{eq:eps3} on $\bepsilon_3$ gives the following bound of the bias:
\begin{equation}
\left\vert\bmu[n]\right\vert \leq  c^{(1)}\sigma^2 + \dfrac1K \left( \dfrac{c_2^{(2)}}{\sigma^2}+c_1^{(2)}+c^{(3)} \right)  + o\left(\dfrac1K\right)\ .
\end{equation}

\subsection{Expression of the Covariance $\bgamma$.}

Outside the measurement interval $I$, denote by $\ell$ the index such that $n=N-1+\ell$. Then, given that $\bh^{(\ell)} = \balpha^{(\ell)} - \balpha_0^{(\ell)}$, we have
\begin{align*}
\bgamma[n,n] &= \left( \balpha_0^{(\ell)}\bz_K\right)^2 + 2\left(\balpha_0^{(\ell)}\bz_K\right)\EE\left\{\bh^{(\ell)}\right\}\bz_K + \EE\left\{\left(\bh^{(\ell)}\bz_K\right)^2\right\} + 2\sigma\balpha_0^{(\ell)}\EE\{\bw_K\bh^{(\ell)}\}\bz_K + 2\sigma\balpha_0^{(\ell)}\bz_K \EE\{\bh^{(\ell)}\bw_K\} \\
& + 2\sigma\EE\{\bh^{(\ell)}\bw_K\bh^{(\ell)}\}\bz_K +\sigma^2\left\|\balpha_0^{(\ell)}\right\|^2 +\sigma^2\EE\left\{\left(\bh^{(\ell)}\bw_K\right)^2\right\}  + 2\sigma^2\balpha_0^{(\ell)}\EE\{\bw_K\bh^{(\ell)}\bw_K\} -\bz[n]^2 - 2\bz[n]\bmu[n]- \bmu[n]^2 \\
&= \sigma^2\left\|\balpha_0^{(\ell)}\right\|^2-\left( \bz[n] - \balpha_0^{(\ell)}\bz_K+\bz[n]\right)^2 + \EE\left\{\left(\bh^{(\ell)}\bz_K\right)^2\right\} + 2\sigma\balpha_0^{(\ell)}\EE\{\bw_K\bh^{(\ell)}\}\bz_K+ 2\sigma\EE\{\bh^{(\ell)}\bw_K\bh^{(\ell)}\}\bz_K \\
&+\sigma^2\EE\left\{\left(\bh^{(\ell)}\bw_K\right)^2\right\}  + 2\sigma^2\balpha_0^{(\ell)}\EE\{\bw_K\bh^{(\ell)}\bw_K\}\\
&\defeq \eta_1[n] + \eta_2[n] + \eta_3[n] + \eta_4[n] + \eta_5[n] + \eta_6[n] + \eta_7[n]\ , 
\end{align*}
where
\begin{align*}
\eta_1[n] &= \sigma^2\left\|\balpha_0^{(\ell)}\right\|^2 \,, & \eta_2[n] &= -\left( \bz[n] - \balpha_0^{(\ell)}\bz_K+\bz[n]\right)^2 \,,
& \eta_3[n] &= \EE\left\{\left(\bh^{(\ell)}\bz_K\right)^2\right\}\,, \\
\eta_4[n] &= 2\sigma\balpha_0^{(\ell)}\EE\{\bw_K\bh^{(\ell)}\}\bz_K\,,
& \eta_5[n] &= \sigma\EE\{\bh^{(\ell)}\bw_K\bh^{(\ell)}\}\bz_K\,,
& \eta_6[n] &= \sigma^2\EE\left\{\left(\bh^{(\ell)}\bw_K\right)^2\right\}\,, \\
\eta_7[n] &= 2\sigma^2\balpha_0^{(\ell)}\EE\{\bw_K\bh^{(\ell)}\bw_K\} \ . 
\end{align*}
Let us now determine an upper bound on each of these terms. 

First, since $\|\balpha_0^{(\ell)}\|^2\leq M\|\bA_0^{\ell}\|_{\max}^2$, we have:
\begin{align}
\nonumber
\eta_1[n] &\leq \sigma^2M\left\|\bA_0^{\ell}\right\|_{\max}^2 \\
\nonumber
&\leq\sigma^2M^{\ell}\left\|\bA_0\right\|^{2\ell}_{\max} \\
&\leq \sigma^2 M^{\ell}\max\left(1,\left(\frac{2J}{M}\right)^{2\ell}\right)\ .
\label{eq:sum1}
\end{align}

Second, by definition of $\epsilon_2$ and $\epsilon_3$ (see expressions~\eqref{eq:eps2.def} and~\eqref{eq:eps3.def}), $\eta_2$ takes the following form:
\begin{equation}
\eta_2[n] = \left(\bepsilon_2[n]+\bepsilon_3[n]\right)^2 = o\left(\dfrac1K\right)\ .
\label{eq:sum2}
\end{equation}

Third, second-order Taylor expansions of $\bh^{(\ell)}$ give:
\begin{align}
\nonumber
\eta_3[n] &\leq
\dfrac{C_\bz^2}{4}\sum_{m,m'=0}^{M-1}\left|\EE\left\{\bh^{(\ell)}[m]\bh^{(\ell)}[m']\right\}\right| \\
\nonumber
&\leq \dfrac{C_\bz^2}{4} \sum_{r,r'=0}^{M(M+1)-1}\left|\left.\dfrac{\partial f_m^{(\ell)}}{\partial \bg[r]}\right\vert_{\bg=\bzero}\left.\dfrac{\partial f_{m'}^{(\ell)}}{\partial \bg[r']}\right\vert_{\bg=\bzero} \EE\{\bg[r]\bg[r']\}\right| + o\left(\dfrac1K\right) \\
\nonumber
&\leq \dfrac{C_\bz^2 M^2(M+1)^2}{4}  \dfrac{d_{1,\bz,M,\ell}^2}{\sigma^4}\dfrac1K \left( C_\bz^2\sigma^2 + 2 \sigma^4\right) + o\left(\dfrac1K\right) \\
&\leq \dfrac{C_\bz^2 M^2(M+1)^2d_{1,\bz,M,\ell}^2}{4K} \left( \dfrac{C_\bz^2}{\sigma^2} + 2\right)  + o\left(\dfrac1K\right)\ .
\label{eq:sum3}
\end{align}

Fourth,
\begin{align*}
\left|\eta_4[n]\right| & \leq \sigma C_\bz\left\|\bA_0^{\ell}\right\|_{\max} \sum_{m,m'=0}^{M-1}\left|\EE\{\bh^{(\ell)}[m]\bw_K[m']\} \right|
\end{align*}
But,
\begin{align}
\nonumber
\left|\EE\{\bh^{(\ell)}[m]\bw_K[m']\} \right|  &\leq \sum_{r=0}^{M(M+1)-1}\left|\left.\dfrac{\partial f_m^{(\ell)}}{\partial \bg[r]}\right\vert_{\bg=\bzero}\right| \left|\EE\{\bg[r]\bw_K[m']\}\right| + o\left(\dfrac1K\right) \\
&\leq M(M+1)\dfrac{d_{1,\bz,M,\ell}}{\sigma^2} C_\bz\dfrac{\sigma}{K} + o\left(\dfrac1K\right)
\label{eq:Ehwm}
\end{align}
Thus,
\begin{equation}
\left|\eta_4[n]\right| \leq M^\ell(M+1)C_\bz^2 d_{1,\bz,M,\ell}\left(\max\left(1,\frac{2J}{M}\right)\right)^\ell\dfrac{1}{K} + o\left(\dfrac1K\right)\ .
\label{eq:sum4}
\end{equation}

Fifth and sixth, second-order Taylor expansions of $\bh^{(\ell)}$ give
\begin{align}
\label{eq:sum5}
\eta_5[n] &= o\left(\dfrac1K\right)\,, \\
\label{eq:sum6}
\eta_6[n] &= o\left(\dfrac1K\right)\ .
\end{align}

Seventh,
\begin{align*}
\left|\eta_7[n] \right| &\leq 2\sigma^2\left\|\bA_0^{\ell}\right\|_{\max} \sum_{m,m'=0}^{M-1} \left| \EE\{\bh^{(\ell)}[m]\bw_K[m]\bw_K[m']\} \right| \\
&\leq 2\sigma^2\left\|\bA_0^{\ell}\right\|_{\max} \sum_{m,m'=0}^{M-1} \sum_{r=0}^{M(M+1)-1}\left|\left.\dfrac{\partial f_m^{(\ell)}}{\partial \bg[r]}\right\vert_{\bg=\bzero}\right| \left| \EE\{\bg^{(\ell)}[r]\bw_K[m]\bw_K[m']\} \right| + o\left( \dfrac1K\right)\ .
\end{align*}
But,
\begin{align*}
\EE\{\bg^{(\ell)}[r]\bw_K[m]\bw_K[m']\} = \sigma\EE\{\bg_1^{(\ell)}[r]\bw_K[m]\bw_K[m']\} + \sigma^2\EE\{\bg_2^{(\ell)}[r]\bw_K[m]\bw_K[m']\}\ .
\end{align*}
As before, since $\EE\{\bg_1^{(\ell)}[r]\bw_K[m]\bw_K[m']\}$ is a third-order moment of a multivariate zero-mean Gaussian vector, it vanishes. And,
\begin{align*}
\left|\EE\{\bg_2^{(\ell)}[r]\bw_K[m]\bw_K[m']\}\right| &= \dfrac1K\left|\sum_{k=0}^{K-1} \EE\left\{\bw_{k}[m_{r}+a_{r}]\bw_{k}[m'_{r}]\bw_K[m]\bw_K[m']\right\} - \delta_{m_r+a_r,m'_r} \EE\left\{\bw_K[m]\bw_K[m']\right\}\right| \\
&= \dfrac1K\left|\sum_{k=0}^{K-1} \delta_{k+m_{r}+a_{r},K+m} \delta_{k+m'_{r},K+m'} +  \delta_{k+m_{r}+a_{r},K+m'} \delta_{k+m'_{r},K+m}\right| \\
&\leq \dfrac2K\ .
\end{align*}
Thus,
\begin{align}
\nonumber
\left| \eta_7[n] \right| &\leq 2\sigma^2\left\|\bA_0^{\ell}\right\|_{\max} M^3(M+1)\dfrac{d_{1,\bz,M,\ell}}{\sigma^2} \dfrac{2\sigma^2}K + o\left( \dfrac1K\right)\\
&\leq 4 M^{\ell+2}(M+1)d_{1,\bz,M,\ell}\left(\max\left(1,\frac{2J}{M}\right)\right)^\ell\dfrac{\sigma^2}{K} + o\left( \dfrac1K\right)\ .
\label{eq:sum7}
\end{align}

To conclude, we combine the expressions~\eqref{eq:sum1},~\eqref{eq:sum2},~\eqref{eq:sum3},~\eqref{eq:sum4},~\eqref{eq:sum5},~\eqref{eq:sum6}, and~\eqref{eq:sum7} to determine an upper bound on the variance $\gamma[n,n]$. It follows:
\begin{align*}
\gamma[n,n] &\leq c_0^{(n)}\sigma^2 + \dfrac{1}{K} \left( \dfrac{c_1^{(n)}}{\sigma^2} + c_2^{(n)} + c_3^{(n)}\sigma^2\right) + o\left(\dfrac1K\right)\ ,
\end{align*}
where
\begin{align*}
c_0^{(n)} & =  M^{\ell}\max\left(1,\left(\frac{2J}{M}\right)^{2\ell}\right) \\
c_1^{(n)} & = \dfrac14C_\bz^4 M^2(M+1)^2d_{1,\bz,M,\ell}^2  \\
c_2^{(n)} & = \dfrac12C_\bz^2 M^2(M+1)^2d_{1,\bz,M,\ell}^2 + M^\ell(M+1)C_\bz^2 d_{1,\bz,M,\ell}\left(\max\left(1,\frac{2J}{M}\right)\right)^\ell\\
c_3^{(n)} & =  4 M^{\ell+2}(M+1)d_{1,\bz,M,\ell}\left(\max\left(1,\frac{2J}{M}\right)\right)^\ell\ .
\end{align*}

\paragraph{If $n'\geq N$}
When $n\geq N$ and $n'\geq N$, applying the Cauchy-Schwarz inequality, we obtain the following bound on the covariance $\gamma[n,n]$:
\begin{align*}
\left|\gamma[n,n']\right| & \leq  \sqrt{\gamma[n,n]\gamma[n',n']} \\
& \leq   \sqrt{c_0^{(n)}c_0^{(n')}}\sigma^2\sqrt{1 + \dfrac{1}{K\sigma^2} \left( \dfrac1{c_0^{(n)}}\left( \dfrac{c_1^{(n)}}{\sigma^2} + c_2^{(n)} + c_3^{(n)}\sigma^2\right) + \dfrac1{c_0^{(n')}}\left( \dfrac{c_1^{(n')}}{\sigma^2} + c_2^{(n')} + c_3^{(n')}\sigma^2\right)\right) + o\left(\dfrac1K\right)}\ .
\end{align*}
A first-order Taylor expansion of this bound as $K\to\infty$ gives
\begin{align*}
& \leq   \sqrt{c_0^{(n)}c_0^{(n')}}\sigma^2 + \dfrac{1}{2K} \left( \sqrt{\dfrac{c_0^{(n')}}{c_0^{(n)}}}\left( \dfrac{c_1^{(n)}}{\sigma^2} + c_2^{(n)} + c_3^{(n)}\sigma^2\right) + \sqrt{\dfrac{c_0^{(n)}}{c_0^{(n')}}}\left( \dfrac{c_1^{(n')}}{\sigma^2} + c_2^{(n')} + c_3^{(n')}\sigma^2\right)\right) + o\left(\dfrac1K\right) \\
& \leq  c_0^{(n,n')}\sigma^2 + \dfrac{1}{K} \left( \dfrac{c_1^{(n,n')}}{\sigma^2} + c_2^{(n,n')} + c_3^{(n,n')}\sigma^2\right) + o\left(\dfrac1K\right)\ , \\
\end{align*}
where
\begin{align*}
c_0^{(n,n')} & =  \sqrt{c_0^{(n)}c_0^{(n')}}\,, \\
c_p^{(n,n')} & =  \dfrac12\left(c_p^{(n)}\sqrt{\dfrac{c_0^{(n')}}{c_0^{(n)}}} + c_p^{(n')}\sqrt{\dfrac{c_0^{(n)}}{c_0^{(n')}}}\right)\,,\quad\forall p\in\{1,2,3\}\ .\\
\end{align*}

\paragraph{If $n'\in I$}
When $n\geq N$ and $n'\in I$, equation~(17) shows us that:
\begin{align*}
\bgamma[n,n']&=\sigma\EE\{\bw[n']\bh^{(\ell)}\}\bz_K + \sigma^2\EE\{\bw[n']\balpha_0^{(\ell)}\bw_K\} + \sigma^2\EE\{\bw[n']\bh^{(\ell)}\bw_K\} \\
& \defeq \beta_1[n,n'] + \beta_2[n,n'] + \beta_3[n,n']\ ,
\end{align*}
where
\begin{align*}
\beta_1[n,n'] &= \sigma\EE\{\bw[n']\bh^{(\ell)}\}\bz_K\,, \\
\beta_2[n,n'] &= \sigma^2\EE\{\bw[n']\balpha_0^{(\ell)}\bw_K\}\,, \\
\beta_3[n,n'] &= \sigma^2\EE\{\bw[n']\bh^{(\ell)}\bw_K\}\ .
\end{align*}
Besides, thanks to the bound~\eqref{eq:Ehwm} we have:
\begin{align}
\nonumber
\left|\beta_1[n,n']\right| &\leq \sigma\dfrac{C_\bz}{2}\sum_{m=0}^{M-1}\left|\EE\{\bw[n']\bh^{(\ell)}[m]\}\right|\\
\nonumber
&\leq \sigma\dfrac{C_\bz}{2}M^2(M+1)\dfrac{d_{1,\bz,M,\ell}}{\sigma^2} C_\bz\dfrac{\sigma}{K} + o\left(\dfrac1K\right)\\
&\leq  M^2(M+1)\dfrac{C_\bz^2 d_{1,\bz,M,\ell}}{2}\dfrac{1}{K} + o\left(\dfrac1K\right)\ .
\label{eq:beta1}
\end{align}
Besides,
\begin{align}
\nonumber
\left|\beta_2[n,n']\right| &\leq \sigma^2\left\|\bA_0^{\ell}\right\|_{\max}\sum_{m=0}^{M-1}\left|\EE\{\bw[n']\bw_K[m]\}\right| \\
\nonumber
&\leq \sigma^2\left\|\bA_0^{\ell}\right\|_{\max} \\
&\leq \sigma^2M^{\ell-1}\left(\max\left(1,\frac{2J}{M}\right)\right)^{\ell}\ .
\label{eq:beta2}
\end{align}
Besides, identically to the bound~\eqref{eq:sum7} on $\eta_7$, we obtain 
\begin{align}
\nonumber
\left|\beta_3[n,n']\right| &\leq \sigma^2\sum_{m=0}^{M-1}\left|\EE\{\bw[n']\bh^{(\ell)}[m]\bw_K[m]\}\right|\\
\nonumber
&\leq \sigma^2 M^3(M+1)\dfrac{d_{1,\bz,M,\ell}}{\sigma^2} \dfrac{4\sigma^2}K + o\left(\dfrac1K\right) \\
&\leq M^3(M+1)d_{1,\bz,M,\ell} \dfrac{4\sigma^2}K + o\left(\dfrac1K\right) \ .
\label{eq:beta3}
\end{align}
Finally, we combine expressions~\eqref{eq:beta1},~\eqref{eq:beta2}, and~\eqref{eq:beta3} to determine an upper bound on the variance $\gamma[n,n']$. It follows:
\begin{align*}
\left|\gamma[n,n']\right| &\leq b_0^{(n,n')}\sigma^2 + \dfrac{1}{K} \left( b_1^{(n,n')} + b_2^{(n)}\sigma^2\right) + o\left(\dfrac1K\right)\ ,
\end{align*}
where
\begin{align*}
b_0^{(n,n')} & =  M^{\ell-1}\left(\max\left(1,\frac{2J}{M}\right)\right)^{\ell} \\
b_1^{(n,n')} & = M^2(M+1)\dfrac{C_\bz^2 d_{1,\bz,M,\ell}}{2} \\
b_2^{(n,n')} & = 4M^3(M+1)d_{1,\bz,M,\ell}  \ .
\end{align*}

\section{Application to an Electrocardiogram}
\begin{figure}
\includegraphics[width=\textwidth]{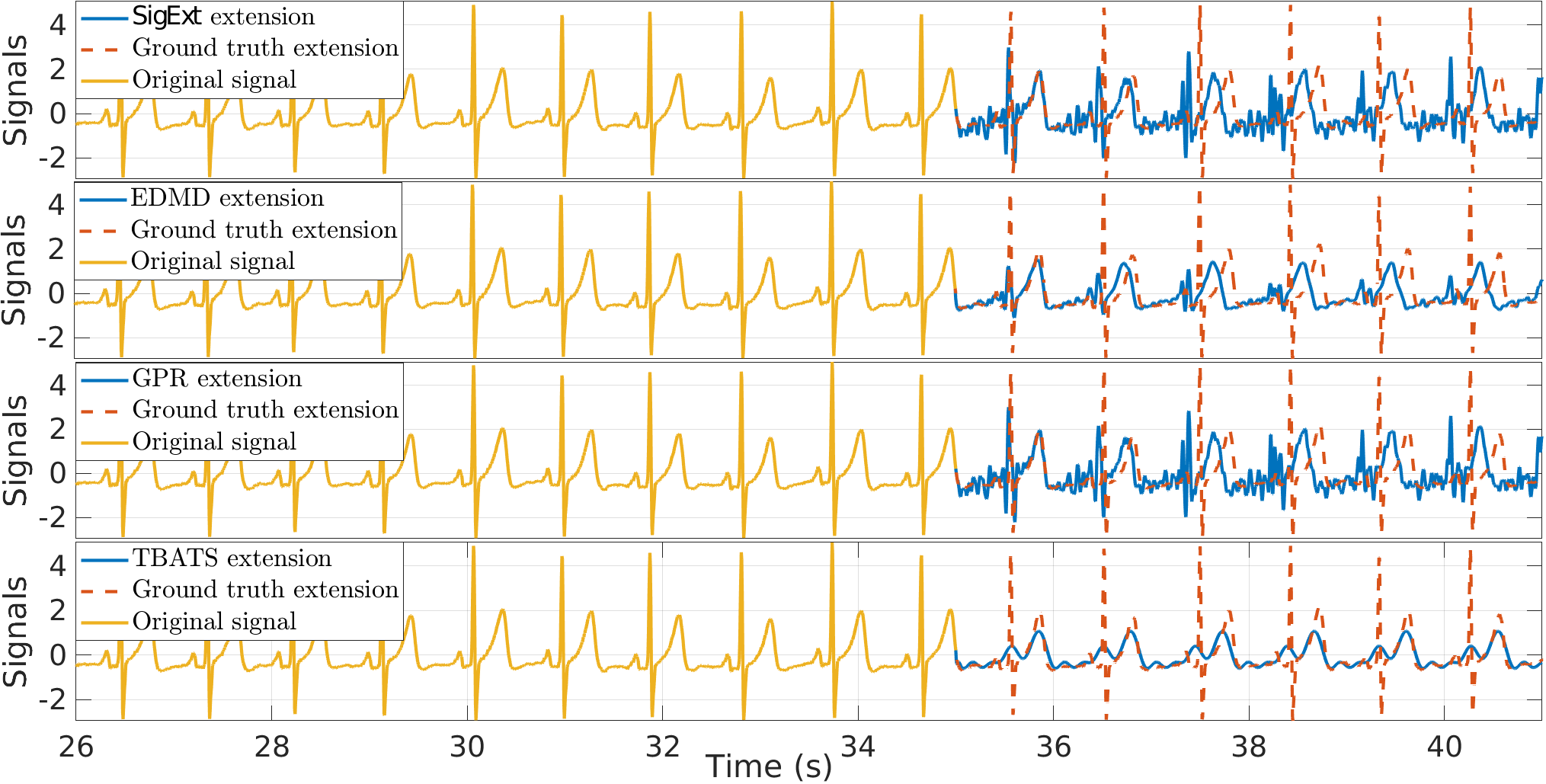}
\caption{Extended ECG (blue) obtained by the {\sf SigExt} forecasting (first panel), the EDMD forecasting (second panel), the GPR forecasting (third panel), and the TBATS forecasting (fourth panel), superimposed with the ground truth signal (dashed red).}
\label{fig:ecg}
\end{figure}

We provide here an additional implementation of {\sf BoundEffRed}, applied to an electrocardiogram (ECG) dataset. The dataset is constructed from a $500$-second-long ECG, sampled at $\fs=200$~Hz, cut into $10$ segments of $50$ seconds each. Fig.~\ref{fig:ecg} depicts the right boundary of one of these subsignals, together with the $6$-second extensions estimated by {\sf SigExt} (first panel), or EDMD (second panel), GPR (third panel), or TBATS (fourth panel). These extensions are superimposed to the ground-truth extension, plotted in red. The sharp and spiky ECG patterns make the AHM model too simplistic to describe this type of signal. Consequently, the forecast produced by {\sf SigExt} is moderately satisfactory. Note that TBATS is the only one that seems to accurately capture the locations of QRS complexes after 37 seconds. We will explore this long-term prediction capability in the future work.

Table~\ref{tab:otd.ecg} contains the median performance index $D$ of the boundary-free TF representations, over the $N$ subsignals evaluated, according to the extension method. As a result of the fair quality of the forecasts, the reduction of boundary effects is less significant than for PPG signal. Nevertheles, the results show that {\sf BoundEffRed} has the same efficiency when the {\sf SigExt} extension, the EDMD extension or the GPR extension is chosen. Indeed, t-tests performed under the null hypothesis that the mean are equals, with a $5\%$ significance level, show no statistical significant difference between {\sf SigExt} and EDMD or GPR, regardless of the representation considered. This justifies the choice of {\sf SigExt} for real-time implementation.

\begin{table}
\centering
\caption{ECG: Performance of the Boundary-Free TF Representations According to the Extension Method}
\begin{tabular}{|c||c|c|c|c|}
  \hline
   \multirow{2}{40pt}{\centering Extension method} & \multicolumn{4}{c|}{Median performance index $D$} \\
   \cline{2-5}
      & STFT & SST & ConceFT & RS\\
   \hhline{|=#=|=|=|=|}
   {\sf SigExt} & $0.584$ & $0.630$ & $0.462$ & $0.642$ \\
   \hline
   Symmetric & $1.199$ & $1.354$ & $1.427$ & $0.943$ \\
   \hline
   EDMD & $0.538$ & $0.558$ & $0.496$ & $0.714$ \\
   \hline
   GPR & $0.639$ & $0.588$ & $0.485$ & $0.616$ \\
   \hline
\end{tabular}
\label{tab:otd.ecg}
\end{table}

\section{Application to a Multicomponent Cardiac Signal}
\begin{figure}
\centering
\includegraphics[width=\textwidth]{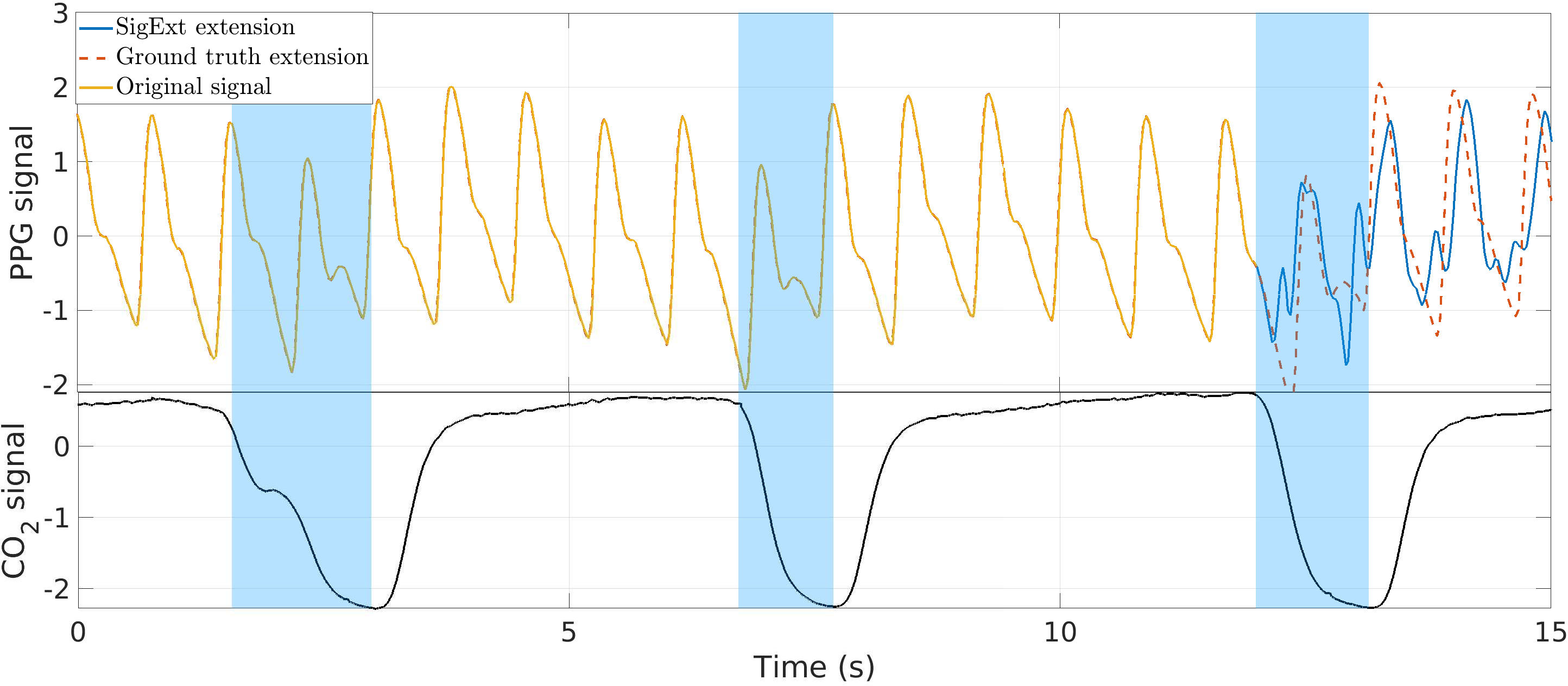}
\caption{PPG signal (top, yellow), and the associated
{\sf SigExt} extension (blue). The simultaneously recorded concentration of $\mathrm{CO}_2$ is below. The bluish areas show the synchrony between the drops in $\mathrm{CO}_2$ concentration and the decreases in the PPG signal.}
\label{fig:ppg2}
\end{figure}

We consider a cardiac signal, namely a photoplethysmogram (PPG) recording, sampled at $300$~Hz. In addition to the cardiac cycle measurement, this signal contains a slow varying component, which is the respiratory component, known as respiration induced intensity variation (RIIV)~\cite{Nilsson13respiration}. The top of Fig.~\ref{fig:ppg2} displays an excerpt of the signal along with a 3-second-long extension obtained by the {\sf SigExt} forecasting. The lower part of Fig.~\ref{fig:ppg2} shows the respiratory signal recording the concentration of $\mathrm{CO}_2$. This signal was recorded simultaneously with the PPG signal, and visually highlights the low-frequency respiratory component contained in the PPG signal. Indeed, the intervals where the concentration of $\mathrm{CO}_2$ drops---highlighted by the bluish areas---coincide with the decreases in the PPG signal. Note also that the forecasting breaks the waveform of the oscillations because of its inability to forecast the high-frequency harmonics contained in the signal. Nevertheless, as long as the low-frequency harmonics of the components contained in the signal are preserved, the forecasting is sufficient to reduce boundary effects in the low-frequency TF domain.

\begin{figure}
\centering
\includegraphics[width=\textwidth]{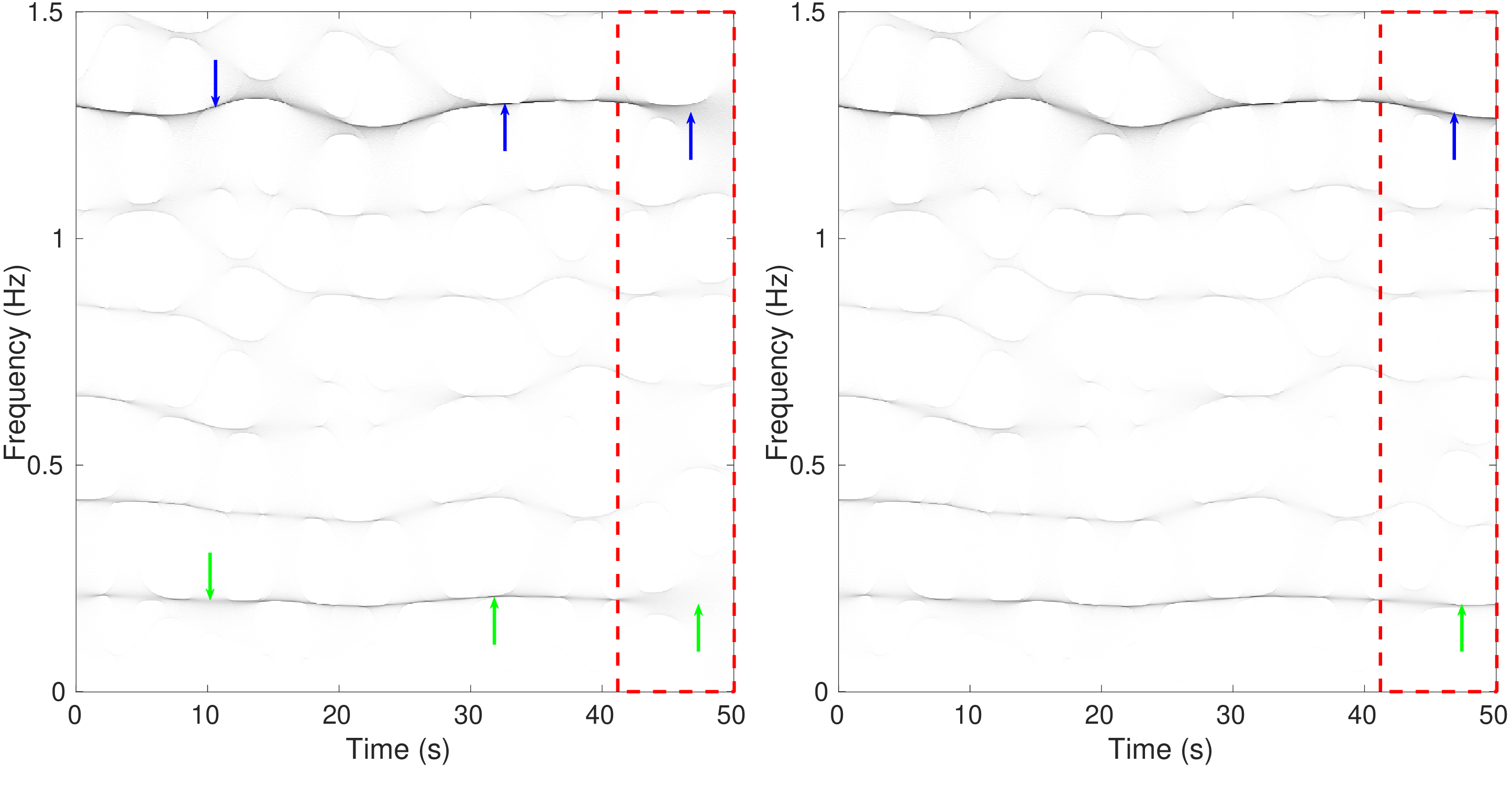}
\caption{Ordinary SST (left) and boundary-free SST (right) of the PPG signal. The window length for the SSTs is 18~seconds. The green arrows indicate the instantaneous frequency of the respiratory component. The blue arrows, on the other hand, indicate the instantaneous frequency of the blood pressure component. The red dashed boxes highlight the areas where boundary effects may appear.}
\label{fig:ppg2.sst}
\end{figure}

The ordinary and boundary-free SSTs of this signal are displayed in Fig.~\ref{fig:ppg2.sst}. These TF analyses brings out both components---the fundamental frequency of the cardiac component, the most energetic one, indicated by the blue arrows, is located around $1.3$~Hz, while the respiratory component, less visible, indicated by the green arrows, is located around $0.2$~Hz  and its multiples. Clearly, near the boundaries, {\sf BoundEffRed} helps improve the quality of the TF representation. Besides, the performance index of this representation takes the value $D=0.491$. This means that {\sf BoundEffRed} has reduced the right-side boundary effects by about $51\%$ with respect to the ordinary SST. This shows the ability of our algorithm to work on signals containing several nonstationary components.

\bibliographystyle{IEEEtran}
\bibliography{../AM041320}

%% file: mydefs.tex
\def\xp{{\mathrm{xp}}}

\def\RR{{\mathbb R}}

\def\EE{{\mathbb E}}

\def\CC{\mathbb C}

\def\NN{\mathbb N}

\def\cN{\mathcal N}
\def\dd#1{\mathrm{d}#1}

\def\1{\mathds{1}}

\def\be{\begin{equation}}
\def\beq#1{\begin{equation}\label{#1}}
\def\ee{\end{equation}}
\def\bea{\begin{eqnarray}}
\def\beqa#1{\begin{eqnarray}\label{#1}}
\def\eea{\end{eqnarray}}
\def\ba{\begin{array}}
\def\ea{\end{array}}

\DeclareMathAlphabet{\mathpzc}{OT1}{pzc}{m}{it}

\def\defeq{\overset{\Delta}{=}}

\def\ie{i.e.\@}
\def\etal{\textit{et al.}}

\def\ccF{{\mathscr F}}

\def\cL{{\mathcal L}}

\def\cN{{\mathcal N}}

\def\ccQ{{\mathscr Q}}
\def\cR{{\mathcal R}}

\def\balpha{\boldsymbol{\alpha}}

\def\bzero{\boldsymbol{0}}

\def\bepsilon{\boldsymbol{\epsilon}}
\def\bgamma{\boldsymbol{\gamma}}

\def\bmu{\boldsymbol{\mu}}

\def\ba{{\mathbf a}}
\def\bA{{\mathbf A}}

\def\be{{\mathbf e}}

\def\bF{{\mathbf F}}

\def\bI{{\mathbf I}}

\def\bw{{\mathbf w}}

\def\bx{{\mathbf x}}
\def\bX{{\mathbf X}}

\def\bY{{\mathbf Y}}

\def\bz{{\mathbf z}}

\def\fs{f_{\mathsf s}}

%% file: algorithm.tex
As explained above, the algorithm for the reduction of boundary effects on TF representations relies on extending the signal by forecasting it before applying the TF analysis.

We start with the notation. Let $x:\RR\to\RR$ denote a continuous-time signal. In this work, we consider a finite-length discretization of that one. Thus, the sampled signal $\bx$, whose length is denoted by $N$, is such that
\[
\bx[n] = x\left(\frac{n}{\fs}\right)\ ,\quad \forall n\in\{0,\ldots,N-1\}\ , 
\]
where $\fs$ denotes the sampling frequency. 
Let $M$ and $K$ be two positive integers such that $M<K$ and $K+M\leq N$. Then, for all $k\in\{0,\ldots,K-1\}$, we extract from $\bx\in\RR^N$ the subsignal $\bx_k\in\RR^M$ given by:
\begin{equation}
\bx_k = 
\begin{pmatrix}
\bx[N-K-M+k] \\
\vdots \\
\bx[N-K+k-1]
\end{pmatrix}\ .
\label{eq:xk}
\end{equation} 
These subsignals are gathered into the matrix $\bX\in\RR^{M\times K}$ such that:
\begin{equation*}
\bX = 
\begin{pmatrix}
\bx_0 & \cdots & \bx_{K-1}
\end{pmatrix}\ .
\end{equation*}
Notice that these subsignals are overlapping each other. Indeed, $\bx_{k+1}$ is a shifting of $\bx_k$ from one sample. We also consider the matrix $\bY\in\RR^{M\times K}$ given by:
\begin{equation*}
\bY = 
\begin{pmatrix}
\bx_1 & \cdots & \bx_{K}
\end{pmatrix}\ .
\end{equation*}
The boundary effect reduction algorithm is based on manipulating $\bX$ and $\bY$.

The pseudo-code of the proposed real-time algorithm to reduce boundary effects on windowing-based TF representations is shown in Algorithm~\ref{alg:boundary}. We coined the algorithm {\sf BoundEffRed}. Below, we detail the algorithm, particularly the signal extension {\sf SigExt} in Algorithm~\ref{alg:extension}.

\begin{algorithm}
\caption{Tackling boundary effects of a TF representation in real-time. $\bF_\bx = \mbox{\sf BoundEffRed}(\bx,M,K,L,\ccF)$}
\label{alg:boundary}
\begin{algorithmic}
\STATE {\bf Inputs}: $\bx_{N}$, $M$, $K$, $N_0$, $L$, $\ccF$
%\STATE {\bf Initialization}: $\bF_\bx^{(N_0)}=\ccF(\bx_{N_0})$
\STATE \vspace{-2mm}
\WHILE {$N$ increases}
\STATE {\bf Real-time input}: $\bx_N$
\STATE \vspace{-2mm}
\STATE {\bf Forecasting step.}
\STATE \quad\textbullet\ Signal extension: $\tilde\bx_{N+L} = \mbox{\sf SigExt}(\bx_N)$. 
\STATE {\bf Representation estimation step.}
\STATE \quad\textbullet\ Extended representation evaluation: $\ccF(\tilde\bx_{N+L})$.
\STATE \quad\textbullet\ Restriction of $\ccF(\tilde\bx_{N+L})$ to the current time interval (see~\eqref{eq:restriction}) to obtain $\bF_\bx^{(N)}=\ccF^{\mathrm{ext}}(\bx_N)$.
\STATE \vspace{-2mm}
\STATE {\bf Real-time output}: Signal representation $\bF_\bx^{(N)}$
%\STATE $N \leftarrow N+1$
\ENDWHILE
\end{algorithmic}
\end{algorithm}

\subsection{Step 1: Extension by Forecasting}

\paragraph{Dynamical Model} 
Establishing a dynamical model means determining the relation linking $\bY$ to $\bX$; that is, finding a function $f$ so that
\begin{equation}
\bY = f(\bX)\ .
\label{eq:generic.model}
\end{equation}
In a general framework, forecasting means estimating the function $f$ from the observed values taken by the signal, in order to predict its future values. For instance, the dynamic mode decomposition~\cite{Schmid10dynamic,Williams15data} or other more complicated models \cite{Roberts13Gaussian,DeLivera11forecasting,west2006bayesian,vlachas2018data}, allow this by setting additional constraints on the behavior of $f$. We will see, in Section~\ref{se:theoretical}, that considering such a complex dynamic model is not necessary for the study of the oscillatory signals of interest to us. That is why we consider here a naive dynamical model, assuming that we have the following relation:
\begin{equation}
\bY = \bA\bX\ ,
\label{eq:dyn.model}
\end{equation}
where $\bA\in\RR^{M\times M}$. In other words, we adopt a classical strategy in the study of dynamical systems, the {\em linearization of a nonlinear system}, when the system is sufficiently regular. Notice that this linearized dynamical model can be written equivalently according to the subsignals $\bx_k$, as:
\begin{equation}
\bx_{k+1} = \bA\bx_{k},\ ,\forall k\in\{0,\dots,K-1\}\ .
\end{equation}
%
%{\color{blue}
%From the Takens' embedding perspective~\cite{Takens81detecting}, $\bX$ can be viewed as a lag map. Under the manifold assumption of the intrinsic phase space that hosts the dynamics, $f$ describes the dynamics supported on the manifold space, and $\bA$ approximates the dynamics on the manifold.
%}

\paragraph{Forecasting}
Determining the forecasting amounts to estimating the unknown matrix $\bA$. Indeed, let $\tilde\bA$ denotes the estimate of $\bA$. 
We then obtain the forecasting of $\bx_{k+\ell}$ by recursively applying the linear relation~\eqref{eq:dyn.model}:
\begin{equation*}
\tilde\bx_{K+\ell} = \underbrace{\tilde\bA\tilde\bA\cdots\tilde\bA}_{\scriptstyle\ell\mbox{ \footnotesize {times}}}\bx_{K} = \tilde\bA^\ell\bx_{K}\ .
\end{equation*}  
Let $\be_M$ denote the unit vector of length $M$ given by $\be_M = \begin{pmatrix} 0 & \cdots & 0 & 1\end{pmatrix}^T$. Consequently, given definition~\eqref{eq:xk}, the forecasting of the signal at time $\frac{N-1+\ell}{\fs}$ is provided by
\begin{equation}
\label{eq:prediction}
\tilde \bx [N-1+\ell] = \be_M^T\tilde\bx_{K+\ell} = \balpha^{(\ell)}\bx_{K}\ ,
\end{equation}  
where $\balpha^{(\ell)}$ denotes the last row of $\tilde\bA^\ell$, \ie, $\balpha^{(\ell)}=\be_M^T\tilde\bA^\ell$.

\paragraph{Model Estimation} To estimate the matrix $\bA$, we consider the simple but numerically efficient least square estimator. That is, we solve the following problem:
\begin{equation}
\label{eq:ls.pb}
\tilde\bA = \arg\min_{\balpha} \cL(\bA)\ ,
\end{equation}
where the loss function $\cL$ is given by:
\[
\cL(\bA) = \|\bY-\bA\bX\|^2 = \sum_{k=0}^{K-1} \|\bx_{k+1}-\bA\bx_k\|^2.
\]
Therefore, solving the problem~\eqref{eq:ls.pb}, \ie, $\nabla \cL(\tilde\bA)=\bzero$, gives the following estimate $\tilde\bA$ of the dynamical model matrix $\bA$:
\begin{align}
\tilde\bA &= \bY\bX^T(\bX\bX^T)^{-1}\ .
\label{eq:lse}
\end{align}

\begin{remark}
This expression clearly shows that the matrix $\tilde\bA$ takes the following form:
\[
\tilde\bA =
\begin{pmatrix}
0       & 1       & 0      & \cdots & 0      \\
\vdots  & \ddots  & \ddots & \ddots & \vdots  \\
\vdots  &         & \ddots & \ddots & 0  \\
0       & \cdots  & \cdots & 0      & 1  \\
\alpha_1& \cdots  & \cdots & \cdots & \alpha_M  \\
\end{pmatrix}.
\]
Then, except for the row vector $\balpha = \left(\alpha_1 \cdots\alpha_M\right)$, the matrix $\bA$ is fully determined by the dynamical model.
\end{remark}

\paragraph{Signal Extension} Since we are building a real-time algorithm, we consider that only the right ``side'' of the signal has to be extended, the left ``side'' being fixed since it only concerns the past values of the signal. We therefore construct the extended signal $\tilde\bx\in\RR^{N+L}$ concatenating the observed signal $\bx$, and the forward forecasting $\tilde\bx_{\mathrm{fw}}\in\RR^L$. We summarize the extension step in Algorithm~\ref{alg:extension}. %We mention that we could, if necessary, handle the backward estimation using the same strategy as described above, but applying it to the reverse signal $\bx^{\rm r} = \begin{pmatrix} \bx[N-1] & \cdots & \bx[0] \end{pmatrix}^T$.

\begin{algorithm}
\caption{Signal extension. $\tilde\bx = \mbox{\sf SigExt}(\bx,M,K,L)$}
\label{alg:extension}
\begin{algorithmic}
\STATE {\bf Inputs}: $\bx_N$, $M$, $K$, $L$
\STATE \vspace{-2mm}
\STATE {\bf Forward forecasting.}
\STATE \quad\textbullet\ Estimation of the matrix $\tilde\bA$ via equation~\eqref{eq:lse}.
\STATE \quad\textbullet\ Forecasting $\tilde\bx_{\mathrm{fw}}\in\RR^L$ obtained applying equation~\eqref{eq:prediction} with $\ell\in\{1,\ldots,L\}$.
%\STATE 
%\STATE {\bf Backward forecasting.}
%\STATE \quad\textbullet\ Reverse signal $\bx$ to $\bx^{\rm r}$. 
%\STATE \quad\textbullet\ LS estimation of the backward matrix $\tilde\bA_{\rm bw}$ via equation~\eqref{eq:lse} applied to $\bx^{\rm r}$.
%\STATE \quad\textbullet\ Reversed backward forecasting $\tilde\bx_{\rm bw}^{\rm r}$ obtained applying equation~\eqref{eq:prediction} to $\bx^r$ with $\ell\in\{1,\ldots,L\}$.
%\STATE \quad\textbullet\ Reverse $\tilde\bx_{\rm bw}^{\rm r}$ to obtain the estimate $\tilde\bx_{\rm bw}$.
\STATE \vspace{-2mm}
\STATE {\bf Output}: Extended signal $\tilde\bx_{N+L} = \begin{pmatrix} \bx_N & \tilde\bx_{\mathrm{fw}}\end{pmatrix}^T$. %\tilde\bx_{\rm bw}  &
\end{algorithmic}
\end{algorithm}

\subsection{Step 2: Extended Time-Frequency Representation}
\label{sse:extension.TFR}
 
Let $\ccF:\RR^{N}\to\CC^{F\times N}$ generically denotes the TF representation of interest to us, which could be, for instance, STFT, CWT, SST, or RS. Here, $F$ typically denotes the size of the discretization along the frequency axis. Due to the boundary effects, the representation $\ccF(\bx_N)$ shows undesired patterns near its edges. To alleviate the boundary effects, we apply the representation to the estimated extended signal $\tilde\bx$. This strategy moves the boundary effects out of the time interval $I=[0, \frac{N-1}{\fs}]$. Finally, the boundary-effects insensitive representation $\ccF^{\mathrm{ext}}:\RR^{N}\to\CC^{F\times N}$ of $\bx_N$ is given for all $\nu\in\{0,\cdots,F-1\}$, $n\in\{0,\cdots,N-1\}$ by:
\begin{equation}
\ccF^{\mathrm{ext}}(\bx_N)[\nu,n] = \ccF(\tilde \bx_{N+L})[\nu,n]\ .
\label{eq:restriction}
\end{equation}
This amounts to restricting the representation $\ccF(\tilde \bx_{N+L})$ to the current measurement interval of $\bx_N$. For the sake of simplicity, we denote the {\em restriction operator} by $\cR$, where $\cR:\CC^{F\times (N+L)}\to\CC^{F\times N}$. Consequently, we have:
\begin{equation}
\ccF^{\mathrm{ext}}(\bx_N) = \cR\left( \ccF(\tilde \bx_{N+L}) \right) \ .
\label{eq:bound.free.TFR}
\end{equation}
We call $\ccF^{\mathrm{ext}}$ the {\em boundary-free TF representation}.

$\bF_\bx^{(N)}\in\CC^{F\times N}$ is the estimation of the boundary-free TF representation at iteration $N$ in Algorithm~\ref{alg:boundary}. For numerical purposes, and to make the real-time implementation achievable, we do not perform a full re-estimation of $\ccF^{\mathrm{ext}}$ at iteration $N+1$. Instead, the additional knowledge provided by $\bx[N+1]$ only influences the values of the last $L_{\mathrm{win}}$ columns of $\ccF^{\mathrm{ext}}(\bx_{N+1})$, where $L_{\mathrm{win}}$ denotes the window half-length used by the TF representation. Thus, $\bF_\bx^{(N+1)}$ is obtained by the concatenation of the first $N-L_{\mathrm{win}}+ 1$ columns of $\bF_\bx^{(N)}$ with the last $L_{\mathrm{win}}$ columns of $\ccF^{\mathrm{ext}}(\bx_{N+1})$.

%% file: theory.tex
\subsection{Signal Model}
\label{sse:model.sine}
We model the meaningful part of the observed signal as a deterministic multicomponent harmonic signal; that is, a sum of sine waves:
\begin{equation}
\bz[n]=\sum_{j=1}^J\Omega_j\cos\left(2\pi f_j \frac{n}{\fs} + \varphi_j\right)\ ,
\label{eq:sum.sine}
\end{equation}
where $J$ denotes the number of components, $\Omega_j>0$ the amplitude of the $j$th component, $f_j$ its frequency, and $\varphi_j\in[0,2\pi)$ its initial phase.
For the sake of simplicity, we make an additional assumption on the frequencies of each component. We assume that for all $j\in\{1,\dots,J\}$:
\begin{equation}
\exists\, p_j,p_j'\in\NN^*: \quad f_j = \dfrac{p_j}{M}\fs = \dfrac{p'_j}{K}\fs\ .
\end{equation}

In addition, the observed signal is assumed to be corrupted by an additive Gaussian white noise. Therefore, the measured discrete signal $\bx$ is written as:
\begin{equation}
\bx = \bz + \sigma\bw\ ,
\label{eq:model.noise}
\end{equation}
where $\bz$ follows model~\eqref{eq:sum.sine}, $\bw$ is a Gaussian white noise, whose variance is normalized to one, and $\sigma>0$. Clearly, $\sigma^2$ denotes the variance of the additive noise $\sigma\bw$.

\subsection{Forecasting Error}
On the forecasting interval, we decompose the estimated signal $\tilde\bx$ as follows:
\begin{equation}
\tilde \bx[n] = \bz[n] + \bepsilon[n]\ ,
\label{eq:forecasting.error}
\end{equation}
where $\bepsilon$ is the forecasting error. When $n\in I=\{0,\dots,N-1\}$, this error contains only the measurement noise, that is $\bepsilon[n]=\sigma\bw[n]$. Outside the interval $I$, the importance of the forecasting error $\bepsilon$ is also affected by the loss of information resulting from the linearization of the dynamical model we consider in~\eqref{eq:dyn.model}. To evaluate the actual behavior of the forward forecasting error $\bepsilon[n]$ when $n\geq N$, we determine its first two moments.
\begin{enumerate}
\item 
The mean, or estimation bias, is such that: 
\begin{align}
\bmu[n]&\defeq\EE\{\bepsilon[n]\} 
=\EE\{\tilde\bx[n]\} - \bz[n]\ .\label{eq:defn:bias0}
\end{align}
Given the forecasting strategy, we have $\bmu[n]=0$ when $n\in I$ and
\begin{equation}
\bmu[n]= \EE\{\balpha^{(n-N+1)}\}\bz_{K} + \sigma\EE\{\balpha^{(n-N+1)}\bw_{K}\} - \bz[n]
\label{eq:bias0}
\end{equation}
when $n\geq N$.
\item 
The covariance is given by:
\begin{align*}
\bgamma[n,n'] &\defeq \EE\{\left(\bepsilon[n]-\bmu[n]\right)\left(\bepsilon[n']-\bmu[n']\right)\}\\
&= \EE\{\tilde\bx[n]\tilde\bx[n']\}-\bz[n]\bz[n'] -\bmu[n]\bz[n'] \\
&\hspace{15pt} - \bmu[n']\bz[n] -\bmu[n]\bmu[n']\ .
\end{align*}
Thus by definition of the noise, we have $\bgamma[n,n'] =\sigma^2\delta_{n,n'}$ when $(n,n')\in I^2$. When $n\geq N$, let us denote $\ell=n-N+1$. Then, we have two cases.
\begin{enumerate}[label=(\roman*)]
\item If $n'\in I$:
\begin{equation}
\bgamma[n,n'] = \sigma\EE\{\bw[n']\balpha^{(\ell)}\}\bz_K + \sigma^2\EE\{\bw[n']\balpha^{(\ell)}\bw_K\}\ .
\label{eq:cov00}
\end{equation}
\item If $n'=N-1+\lambda\geq N$:
\begin{align}
\nonumber
\hspace{-6pt}\bgamma[n,n'] &= \bz_K^T\EE\left\{{\balpha^{(\ell)}}^T\balpha^{(\lambda)}\right\}\bz_K + \sigma\EE\{\balpha^{(\ell)}\bw_K\balpha^{(\lambda)}\}\bz_K \\
\nonumber
&\hspace{-3pt} + \sigma\EE\{\balpha^{(\lambda)}\bw_K\balpha^{(\ell)}\}\bz_K + \sigma^2\EE\{\balpha^{(\ell)}\bw_K\balpha^{(\lambda)}\bw_K\}  \\
&\hspace{-3pt} -\!\bz[n]\bz[n']-\!\bz[n]\bmu[n']-\!\bz[n]\bmu[n'] -\! \bmu[n]\bmu[n']\,.
\label{eq:cov01}
\end{align}
\end{enumerate}
Besides, we recall that $\bgamma[n,n']=\bgamma[n',n]$.
\end{enumerate}

In Theorem~\ref{th:error}, we specify the asymptotic behavior of the forecasting bias and covariance when the dataset size $K$ is great.
\begin{theorem}
\label{th:error}
Let $\bx\in\RR^N$ be a discrete-time random signal following model~\eqref{eq:model.noise}. Let $\tilde\bx$ denotes its forecasting, obtained using the extension Algorithm~\ref{alg:extension}. Let $n\geq N$ be a sample index. Then, the first-order moment of the forecasting error $\epsilon[n]$ in~\eqref{eq:forecasting.error} defined in \eqref{eq:defn:bias0} satisfies
\begin{equation}
\left|\bmu[n]\right| \leq a^{(n)}_0\sigma^2 + \dfrac1K \left( \dfrac{a^{(n)}_1}{\sigma^2} + a^{(n)}_2 \right) + o\left( \dfrac1K \right)
\label{eq:mean.error}
\end{equation}
as $K\to\infty$, where $a^{(n)}_0$, $a^{(n)}_1$, and $a^{(n)}_2$ are positive quantities, independent of $K$ and $\sigma$.
Its second-order moment $\bgamma[n,n']$ satisfies the following approximation equations:
\begin{enumerate}[label=(\roman*)]
\item if $n'\in I=\{0,\ldots,N-1\}$:
\begin{equation}
\left|\bgamma[n,n']\right|\leq b^{(n,n')}_0\sigma^2 + \!\dfrac1K\! \left( b^{(n,n')}_1\! + b^{(n,n')}_2\sigma^2 \! \right) + o\left( \dfrac1K \right)
\label{eq:cov.error.1}
\end{equation}
as $K\to\infty$, where $b^{(n,n')}_0$, $b^{(n,n')}_1$, and $b^{(n,n')}_2$ are positive quantities, independent of $K$ and $\sigma$;
\item if $n'\geq N$:
\begin{align}
\nonumber
\left|\bgamma[n,n']\right| &\leq c^{(n,n')}_0\sigma^2 \\
&\hspace{10pt} + \!\dfrac1K\! \left(\! \dfrac{c^{(n,n')}_1}{\sigma^2} + c^{(n,n')}_2 + c^{(n,n')}_3\sigma^2 \right) + o\!\left( \!\dfrac1K \!\right)
\label{eq:cov.error.2}
\end{align}
as $K\to\infty$, where $c^{(n,n')}_0$, $c^{(n,n')}_1$, $c^{(n,n')}_2$ and $c^{(n,n')}_3$ are positive quantities, independent of $K$ and $\sigma$.
\end{enumerate}
\end{theorem}

\begin{proof}
See the Supplementary Materials. The proof is mainly based on Taylor expansions of the forecasting error. Nonoptimal values of $a^{(n)}_0$, $a^{(n)}_1$,\dots, $b^{(n,n')}_0$,\dots, $c^{(n,n')}_3$ are also provided in the proof. 
\end{proof}
%The Isserlis' theorem~\cite{Isserlis16formula}, which provides a formula for the computation of higher-order moments of Gaussian random variables.

The forecasting bias and covariance asymptotically depend linearly on the ratio $\frac{1}{K}$. This shows the need to use a sufficiently large dataset to obtain an accurate forecast. Ideally when $K\to\infty$, the forecasting error would behave like the measurement noise $\sigma\bw$, \ie, a zero-mean white noise whose variance is of the order of $\sigma^2$. However, Theorem~\ref{th:error} shows that the estimator may remain asymptotically biased (the first term in the bound of equation~\eqref{eq:mean.error} being independent of $K$). Concerning the covariance of the forecasting error, the expected asymptotic behavior is verified. Indeed, when $K \to\infty$, the variance of the forecasting estimator increases at most linearly with the noise variance $\sigma^2$, since is bounded by $b_0^{(n,n')}\sigma^2$.

Assume now that $K$ is sufficiently great and fixed. The noise influences the quality of the forecasting estimator in two ways. On the one hand, when the noise variance $\sigma^2$ is great, the bias and the variance of the estimator potentially increase linearly with $\sigma^2$. This reflects the classical behavior of an estimator of a signal degraded by additive noise. On the other hand, due to terms in $1/\sigma^2$, when the noise variance $\sigma^2$ is small, the bias and the variance of the forecasting estimator are no longer controlled via equations~\eqref{eq:mean.error} and~\eqref{eq:cov.error.2}. This illustrates the necessary presence of noise to obtain an accurate prediction. Indeed, if $\sigma$ is too small, the matrix $\bX\bX^T$ is ill-conditioned and its inversion in~\eqref{eq:lse} is sensitive. The forecasting is then of poor quality. Noise therefore appears as a regularization term to ensure the invertibility of $\bX\bX^T$. This seemingly counterintuitive fact indicates the challenge we encounter to do forecasting with the dynamic model~\eqref{eq:generic.model}, the degeneracy. Note that from the lag map perspective, $\bX$ and $\bY$ consist of points located on a low dimensional manifold. Thus, locally the ranks of $\bX$ and $\bY$ are low, which leads to the degeneracy. However, when noise exists, it naturally fills in deficient ranks, and stabilizes the algorithm.

The dependencies of the forecasting quality on the subsignals lengths $M$ and the forecasting index $\ell=n-N+1$ are hidden in the expression of the parameters $a^{(n)}_0$, $a^{(n)}_1$,\dots, $b^{(n,n')}_0$,\dots, $c^{(n,n')}_3$. The numerical results, discussed in Section~\ref{ssse:res.sine}, allow us to better evaluate these dependencies.

\subsection{Adaptive Harmonic Model}
\label{RemarkAHM}

One can extend the previous result to the case where the instantaneous frequencies and amplitudes of the components of the deterministic part of the observed signal are {\em slowly varying}. It is an \textit{AM-FM model} that, in its is continuous-time version, takes the following form
\begin{equation}
z(t) = \sum_{j=1}^J a_j(t)\cos(2\pi\phi_j(t))\,,
\end{equation}
where $a_j(t)$ and $\phi'_j(t)$ describe how large and fast the signal oscillates at time $t$. 
Clearly, \eqref{eq:sum.sine} is a special case satisfying the AM-FM model when the $a_j$ and $\phi'_j$ are both constants. 
In many practical signals, the amplitude and frequency do not change fast. It is thus reasonable to further restrict the regularity and variations of the instantaneous amplitudes and frequencies of these AM-FM functions. When the speeds of variation of the instantaneous amplitudes $a_j$ and frequencies $\phi'_j$ are small, the signal can be ``locally'' well approximated by a harmonic function in \eqref{eq:sum.sine}; that is, 
locally at time $t_0$, $z(t)$ can be well approximated by $z_0(t) := \sum_{j=1}^J a_j(t_0)\cos(2\pi(\phi_j(t_0)-t_0\phi'_j(t_0)+\phi'_j(t_0)t))$ by the Taylor expansion. An AM-FM function satisfying the slow variation of the instantaneous amplitudes $a_j$ and frequencies $\phi'_j$ is referred to the adaptive harmonic model (AHM) (see~\cite{Chen14nonparametric,Daubechies16conceft} for mathematical details).
It is thus clear that the forecasting error caused by the {\sf SigExt} algorithm additionally depends on the speed of variation of the instantaneous amplitudes $a_j$ and frequencies $\phi'_j$. 
For the forecasting purpose, it is thus clear that when the speed of variation of $a_j$ and $\phi'_j$ are slow $K$ can be chosen large while the signal can be well approximated by~\eqref{eq:sum.sine}. Hence, Theorem~\ref{th:error} can still be applied to approximate the error. However, how to determine the optimal $K$ based on the signal depends on the application and is out of the scope of this paper. It will be explored in future work for specific application problems.
%\end{remark}

The models described in Sections~\ref{sse:model.sine} and~\ref{RemarkAHM} consider the meaningful part of the signal to be a deterministic component. These models are purposely adapted to signals showing local line spectra. The physiological signals we are interested in, such as respiratory or cardiac signals, have this characteristic (see Section~\ref{sse:physio.sig}). Signals with wider local spectra, such as electroencephalography signals, do not fall into this category, and are more faithfully modeled as random signals. Thus, the theoretical justifications proposed above are no longer applicable to guarantee the forecasting quality of {\sf SigExt}.

\subsection{Performance of the Boundary Effects Reduction}
\label{sse:perf.BoundEffRed}
While we do not provide a generic proof of the boundary effect reduction on {\em any} TF analysis tools, we discuss a particular case of SST. Since SST is designed to analyze signals satisfying the AHM, as is discussed in Section~\ref{RemarkAHM}, Theorem~\ref{th:error} ensures that the forecasting error $\bepsilon$ defined in~\eqref{eq:forecasting.error} is controlled and bounded in terms of mean and covariance. Recall Theorem 3 in~\cite{Chen14nonparametric}, which states that when the additive noise is stationary and may be modulated by a smooth function with a small fluctuation, the SST of the observed signal remains close to the SST of the clean signal, throughout the TF plane. 
We refer the reader to~\cite{Chen14nonparametric} for a precise quantification of the error made in the TF plane, which depends notably on the covariance of the additive noise and on the speed of variation of the amplitudes and instantaneous frequencies composing the signal. 
In our case, while the noise of the historical data is stationary, the forecasting error depends on the historical noise, and hence the overall ``noise'' is nonstationary. However, the dependence only appears in the extended part of the signal. We claim that the proof of Theorem 3 in~\cite{Chen14nonparametric} can be generalized to explain the robustness of SST to the noise, and a systematic proof will be reported in our future theoretical work. We verify experimentally that Algorithm~\ref{alg:boundary} is efficient for a large number of representations in the following Section~\ref{se:results}. 
Therefore, in our case, this means that the boundary effect is strongly reduced since the impact of the forecasting error does SST is limited. An immediate application is that the instantaneous frequencies can now be well estimated continuously up to the edges of the TF plane, and real-time acquisition of the instantaneous frequency and amplitude information is possible.

%% file: results.tex
For the reproducibility purpose, the MATLAB code and datasets used to produce the numerical results in this section are available online at \url{https://github.com/AdMeynard/BoundaryEffectsReduction}.

\subsection{Extension Methods for Comparison}
\label{sse:methods}
To highlight the fact that the linear dynamical model is sufficient to catch most of the dynamical behavior of signals following the AHM, we compare the performance of Algorithm~\ref{alg:boundary} with a simple extension obtained by pointwise symmetrization~\cite{Kharitonenko02wavelet}. We also evaluate the performance of reference forecasting algorithms that could be used for extending such signals. These methods are:
\begin{itemize}
\item The EDMD has been developed by Williams \etal~\cite{Williams15data}. The proposed algorithm is a way to obtain an approximation of the so-called Koopman operator of the observed system, which theoretically allows catching dynamic of nonlinear systems~\cite{Korda18linear}.
\item The GPR~\cite{Rasmussen06gaussian} is a method relying on a probabilistic dynamical model. That one is based on the Gaussian process structure, and therefore offers more flexibility in the type of dynamic that could be modeled than the linear model~\eqref{eq:dyn.model}.
\item The TBATS method~\cite{DeLivera11forecasting} is based on a classical decomposition of times series into a trend, seasonal and ARMA components, with a specific dynamic for the seasonal component. This model demands the estimation of numerous parameters and, by implication, may require a long computation time.
\end{itemize}

\subsection{Evaluation Metrics}

The first evaluation metric is quantifying the forecasting quality (\ie~not depending on $\ell$) in the time domain. We evaluate the Experimental Mean Square Error $\mathrm{MSE_{xp}}(\tilde\bx)$ of the forward forecast extended signals, namely,
\begin{align}
\label{eq:mse}
\mathrm{MSE_{\xp}}(\tilde\bx) &= \dfrac1{L}\|\tilde\bx -\bx^\mathrm{ext}\|^2 \\[-1mm]
\nonumber
&=\! \dfrac1{L}\sum_{\ell=1}^L \bmu_{\xp}[N\!-\!1\!+\!\ell]^2 \!+\! \bgamma_{\xp}[N\!-\!1\!+\!\ell,N\!-\!1\!+\!\ell] .
\end{align}
where $\bx^\mathrm{ext}$ is the {\em ground-truth extended} signal; that is, $\bx^\mathrm{ext} := \begin{pmatrix}\bx[-L] & \cdots & \bx[N-1+L] \end{pmatrix}$. Then, as long as the bias $\bmu[N-1+\ell]$ and the variance $\bgamma[N-1+\ell,N-1+\ell]$ of the forecasting estimator remain small for all $\ell$, the MSE takes small values either.

The second evaluation metric is evaluating the quality of the boundary effect reduction directly on the TF representation. We compare the obtained TF representation to the optimal TF representation $\ccF_N^\mathrm{opt}(\bx)$, defined as the restriction of the representation of the ground-truth extended signal $\bx^\mathrm{ext}$; that is, we have
\begin{equation}
\ccF^\mathrm{opt}(\bx_N) = \cR\left( \ccF(\bx^\mathrm{ext}) \right) \, ,
\label{eq:opt.TFR}
\end{equation} 
where $\mathcal{R}$ is defined in \eqref{eq:bound.free.TFR}.
To compare different techniques, we use a criterion proposed in~\cite{Daubechies16conceft} that quantifies the optimal transport (OT) distance between a given TF representation and the optimal one. In general, the OT distance is a quantity that measures how different two probability density functions are. Denote a TF representation as $\ccQ$. For $t$ fixed, we consider the probability density function defined by $p_\ccQ^t(\xi) = \left. |\ccQ(\xi,t)|^2 \middle/ \int_\RR |\ccQ(\nu,t)|^2\dd\nu \right.$; that is, normalizing the spectral information at time $t$. At each time $t$, the OT distance $d_{t}$ between two TF representations $\ccQ$ and $\ccF^\mathrm{opt}$ at time $t$ is defined by
\begin{equation*}
d_{t}(\ccQ,\ccF^\mathrm{opt}) = \int_\RR\left|\tilde P_{\ccQ}^t(\xi)-  P_{\ccF^\mathrm{opt}}^t(\xi)\right|\dd\xi\ ,
\end{equation*}
where $P_{\ccQ}^t(\xi)=\int_{-\infty}^\xi p_{\ccQ}^t(\nu)\dd\nu$ and $\tilde P_{\ccF^\mathrm{opt}}^t(\xi)=\int_{-\infty}^\xi\tilde p_{\ccF^\mathrm{opt}}^t(\nu)\dd\nu$. In other words, the OT distance between $\ccQ$ and $\ccF^\mathrm{opt}$ at time $t$ is given by the $L^1$ norm of the associated cumulative density function.  %
Note that $d_{t}(\ccQ,\ccF^\mathrm{opt})$ quantifies the proximity between the estimated and actual TF representations at time $t$. With the OT distance, we define the {\em performance index} $D(\ccQ,\ccF)$ for the reduction of boundary effects of a given TF representation $\ccQ$ over another TF representation $\ccF$ by
%\begin{equation}
%D(\ccQ) = \dfrac1{|I|}\bigintssss_I\ \frac{d_{t}\left(\ccQ,\ccF^\mathrm{opt}\right)}{d_{t}\left(\ccF,\ccF^\mathrm{opt}\right)}\ \dd t\ .
%\label{eq:index.perf}
%\end{equation}
\begin{equation}
D(\ccQ,\ccF) = \dfrac{\displaystyle\int_I d_{t}\left(\ccQ,\ccF^\mathrm{opt}\right)\dd t}{\displaystyle\int_I d_{t}\left(\ccF,\ccF^\mathrm{opt}\right)\dd t}\ .
\label{eq:index.perf}
\end{equation}
In other words, $D(\ccQ,\ccF)$ is the ratio between its averaged OTD to the optimal TF representation $\ccF^\mathrm{opt}$ (see~\eqref{eq:opt.TFR}) and the averaged OTD of the original TF representation $\ccF$ to $\ccF^\mathrm{opt}$.
Thus, $D(\ccQ)<1$ means a reduction of the boundary effects. Let us evaluate the quality of the boundary effects reduction on biomedical signals.

\subsection{Simulated Signals}
We first evaluate the quality of the forecasting step and compare it to the theoretical results provided by Theorem~\ref{th:error}. The level of the forecasting error depends on at least two parameters:
\begin{itemize}
\item The noise variance $\sigma^2$.
\item The size of the training dataset $K$. 
\end{itemize}
In Sections~\ref{ssse:res.sine} and~\ref{ssse:res.ahm}, we study the influence of these parameters. A comparison with the theoretical results of Section~\ref{se:theoretical} is also available.

\subsubsection{Sum of sine waves}
\label{ssse:res.sine}
We proved that the linear dynamic model is sufficient to catch the dynamical behavior of signals taking the form~\eqref{eq:sum.sine}. In order to validate this theoretical result, we apply the forecasting Algorithm~\ref{alg:extension} to a large number of realizations of the random vector $\bx$ of size $N=10^4$, following model~\eqref{eq:model.noise}, and such that the deterministic component $\bz$ takes the form:
\[
\bz[n]\! =\!\cos\!\left(2\pi p_1 \dfrac{n}{M} \right)\! +\! A\cos\!\left(2\pi p_2 \dfrac{n}{M} \right),\quad \!\!\forall n\!\in\!\{1,\ldots,N\},
\]
with $M=150$, $p_1=10$, $p_2=33$ and $A=1.4$. Besides, the additive noise is chosen to be Gaussian: $\bw\sim\cN(\bzero,\bI)$.

\begin{figure}
\centering
\includegraphics[width=.48\textwidth]{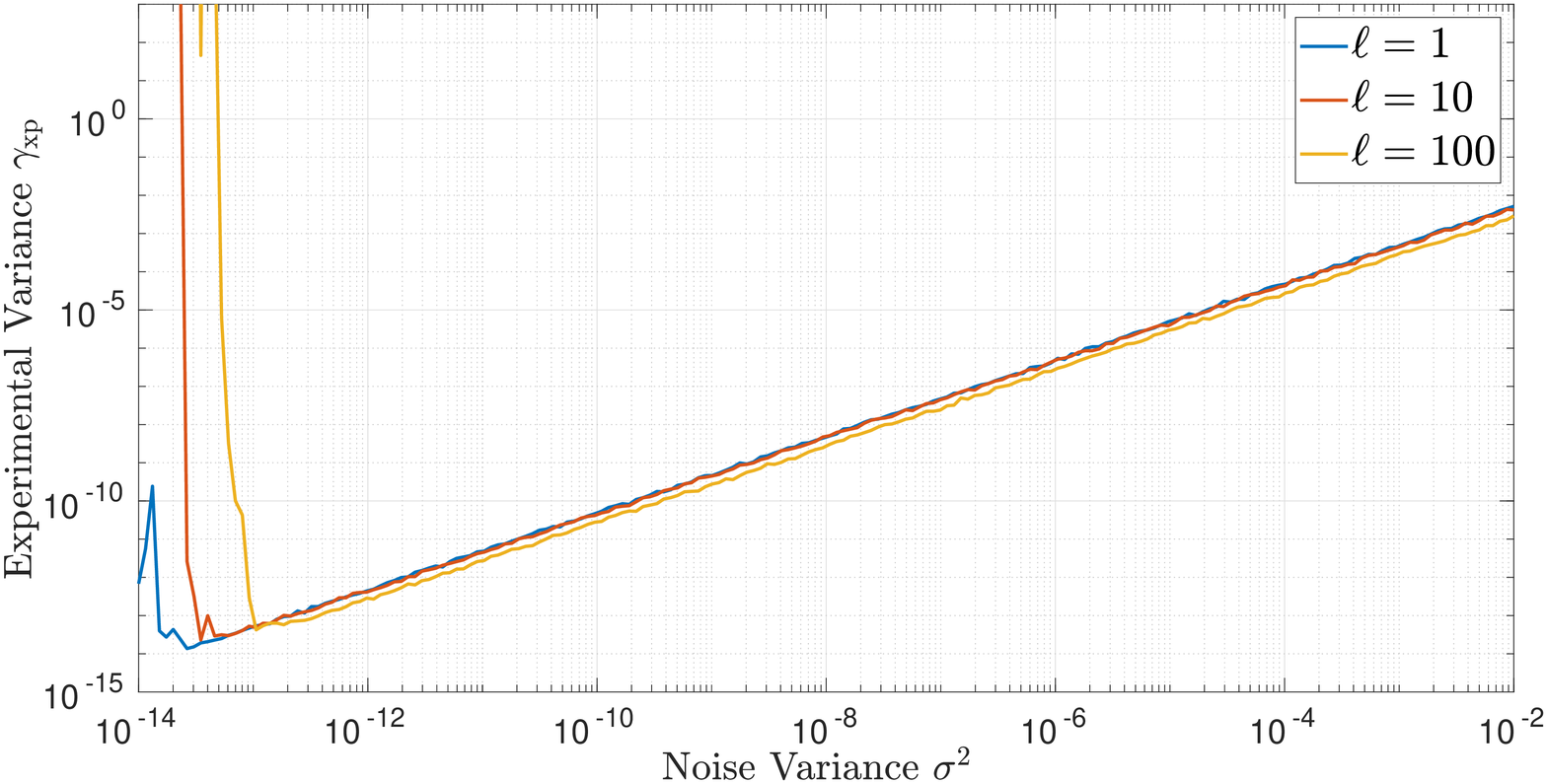}
\caption{Evolution of the experimental forecasting variance as a function of the noise variance for three different values the forecasting sample index $\ell$.}
\label{fig:res.noise.sine}
\end{figure}

\paragraph{Influence of the Noise Variance $\sigma^2$} Here, the size of the training dataset is set to $K=450$. Then, the forecasting algorithm is run on $500$ realizations on the discrete signal $\bx$ for $200$ different values of $\sigma$, logarithmically equispaced from $10^{-7}$ to $10^{-1}$. For each of these values, we determine the experimental bias, denoted as $\mu_{\xp}[N-1+\ell]$, and experimental variance, denoted as $\gamma_{\xp}[N-1+\ell,N-1+\ell]$. Fig.~\ref{fig:res.noise.sine} shows the experimental forecasting variance as a function of the noise variance for three different values the forecasting sample index: $\ell=1$, $\ell=10$, and $\ell=100$.

First, when the noise is weak, the variance of the forecasting estimator is dramatically high. For instance, in this specific example, the estimator $\tilde\bx[N-1+\ell]$ for $\ell=10$ is inaccurate as soon as the noise variance $\sigma^2$ is lower than $4\times 10^{-14}$ (see the red line in Fig.~\ref{fig:res.noise.sine}). This validates the result of Theorem~\ref{th:error}, highlighting the necessary presence of noise to ensure the numerical stability of the estimator. Furthermore, as soon as $\sigma^2$ is higher than this critical threshold, the estimator variance increases linearly with $\sigma^2$, as predicted by the bound~\eqref{eq:cov.error.2} provided in Theorem~\ref{th:error}. Note that the parameter $\ell$ has little influence on the quality of the estimator. This is due to the \textit{stationarity} of the studied signal, allowing long-range forecasts without deterioration of the estimator.

\begin{figure}
\includegraphics[width=.48\textwidth]{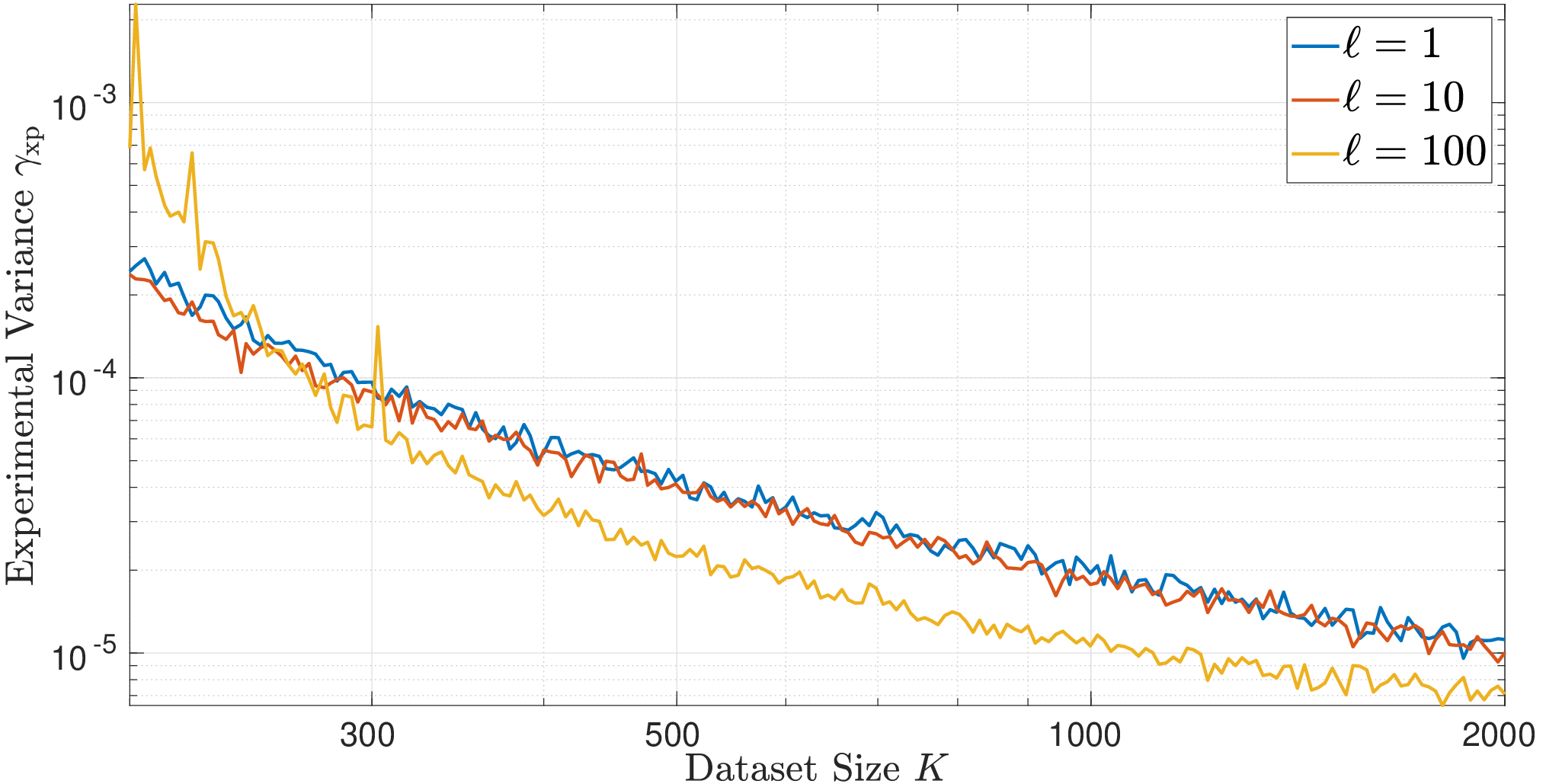}
\caption{Evolution of the experimental forecasting variance in function of the dataset size for three different values the forecasting sample index $\ell$.}
\label{fig:res.size.sine}
\end{figure}

\paragraph{Influence of the Training Dataset Size $K$} Here, the noise variance $\sigma$ is set to $\sigma=10^{-2}$. Then, {\sf SigExt} is run on $500$ realizations of the discrete signal $\bx$ for $200$ different values of training dataset size $K$, logarithmically equispaced from $450$ to $2000$. For each of these values, we determine the experimental bias $\mu_{\xp}[N-1+\ell]$ and variance $\gamma_{\xp}[N-1+\ell,N-1+\ell]$. Fig.~\ref{fig:res.size.sine} shows the experimental forecasting variance as a function of the training dataset size for three different values the forecasting sample index: $\ell=1$, $\ell=10$, and $\ell=100$.

As anticipated by Theorem~\ref{th:error}, the variance of the estimator decreases linearly with $\frac1K$ as long as the dataset size is large enough. Indeed, in the log-log plot displayed in Fig.~\ref{fig:res.size.sine}, as soon as $K>700$, each of the three curves representing the forecasting variance as a function of $K$ tends towards an asymptote of slope $-1$. 
In practice, the asymptotic behavior of {\sf SigExt}, illustrated here on a synthetic signal, can be generalized to the study of real-life signals, as long as the duration of the recorded signal largely exceeds that of the desired extension, which allows the user to set $K\gg L$.

This study overlooks the analysis of the influence of the parameter $M$, whose influence on the value of the experimental variance is numerically not significant as long as $M\ll 2K$ and $\ell<M$. The choice of this parameter is especially crucial when the deterministic component of the signal is no longer stationary. The AHM, discussed below, is an example.

\subsubsection{Adaptive Harmonic Model}
\label{ssse:res.ahm}
Consider a signal satisfying the AHM so that the instantaneous frequencies and amplitudes of its components vary over time. The deterministic component $\bz$ of the random vector $\bx$, constructed following the model~\eqref{eq:model.noise}, takes the following form, for all $n\in\{1,\ldots,N\}$:
\[
\bx[n] = \cos\left(2\pi \phi_1[n] \right) + R[n]\cos\left(2\pi \phi_2[n] \right) \ ,
\] 
where the instantaneous amplitude $R$ is given by:
\[
R[n] = 1.4 + 0.2\cos\left(4\pi\frac{n}{N}\right)\ ,
\]
and the instantaneous phases are such that:
\begin{align*}
\phi_1[n] &= \frac{p_1}{P}\left( n + \frac{0.01}{2\pi}\cos\left(2\pi\frac{n}{N}\right) \right) \\[-1mm]
\phi_2[n] & = p_2\frac{n}{P} + \frac{20}{2N\fs}n^2\ .
\end{align*}
Besides, the noise is chosen to be Gaussian, $\bw\sim\cN(\bzero,\bI)$, and we take $N=10^4$, $P=750$, $p_1=10$, $p_2=23$.

\paragraph{Influence of the subsignals length $M$}
To evaluate the forecasting quality, {\sf SigExt} is applied to $1000$ realizations of the above-described signal. We forecast signal extensions of $0.1$~s, \ie, $L=700$. Table~\ref{tab:mse.sine} shows the resulting mean and standard deviation (SD) of the experimental MSE, according to three different values of the subsignals length $M$. Note that choosing too small an $M$ does not provide enough information to forecast the signal satisfactorily. Furthermore, too large an $M$ makes the algorithm insensitive to local nonstationarities produced by frequency variations. The user must then make a compromise on the choice of $M$ in order to optimize the performance of the algorithm. A choice of $M$ such that the subsignals contain about a dozen oscillations is generally appropriate.

\paragraph{Comparison of the extensions methods}
The same set of experiments is run on the extension methods outlined in Section~\ref{sse:methods}. The results, displayed in Table~\ref{tab:mse.sine}, show that the naive extension we propose gives satisfying results. Indeed, when the subsignal length $M$ is optimally chosen, {\sf SigExt} performance is of the same order as that of ore sophisticated methods, like GPR or TBATS. Besides, even though these methods could be slightly more robust to perturbations, they are substantially limited by the computing time they require, which prevents them from being used to exploit real-time data. Thus, {\sf SigExt} is the extension method that optimizes the trade-off between forecasting quality and computing time.
Besides, the last column of Table~\ref{tab:mse.sine} gives the performance index $D$ of the boundary-free STFT associated with each extension method. This illustrates the ability of our algorithm to reduce boundary effects to a level comparable with what is possible with the GPR or TBATS extensions. The performance of {\sf BoundEffRed} on other TF representations is more thoroughly analyzed on real-life signals in the following.

\begin{table}
\centering
\caption{AHM Signal: Performance of the Extension Methods and the Associated Boundary-Free STFT}
\begin{tabular}{|c||c|c||c|}
  \hline
   \multirow{2}{60pt}{\centering Extension method}  & \multirow{2}{30pt}{\centering CPU time (s)} & \multirow{2}{50pt}{\centering MSE\\ (mean $\pm$ SD)} & \multirow{2}{50pt}{\centering Index $D$ of the STFT}\\
    &  & & \\
   \hhline{|=#=|=#=|}
   {\sf SigExt ($M=100$)} & $0.008$ & $1.133 \pm  0.077$ & $0.0091 \pm  0.0019$\\
   \hline
   {\sf SigExt ($M=750$)} & $0.122$ & $0.479 \pm  0.166$ & $0.0056 \pm  0.0019$\\
   \hline
   {\sf SigExt ($M\!=\!1500$)} & $0.634$ & $0.907 \pm  0.377$ & $0.0065 \pm  0.0023$\\
   \hline
%   {\sf SigExt ($M\!=\!3000$)} & $2.881$ & $5.365 \pm  3.919$ & $0.0117 \pm  0.0038$\\
%   \hline
   Symmetric &  $<0.001$ & $5.452 \pm  0.002$ & $1.8209 \pm  0.0025$\\
   \hline
   EDMD & $1.587$ & $0.983  \pm  0.004$ & $0.1145 \pm  0.0018$\\
   \hline
   GPR & $515.787$ & $0.479  \pm  0.166$ & $0.0119 \pm  0.0037$\\
   \hline
   TBATS  & $1776.778$ & $0.686 \pm  0.467$ & $0.3847 \pm  0.1397$\\
   \hline
\end{tabular}
\label{tab:mse.sine}
\end{table}

\subsection{Real Physiological Signals}
\label{sse:physio.sig}

\subsubsection{Respiratory Signal}
\begin{figure}
\includegraphics[width=.48\textwidth]{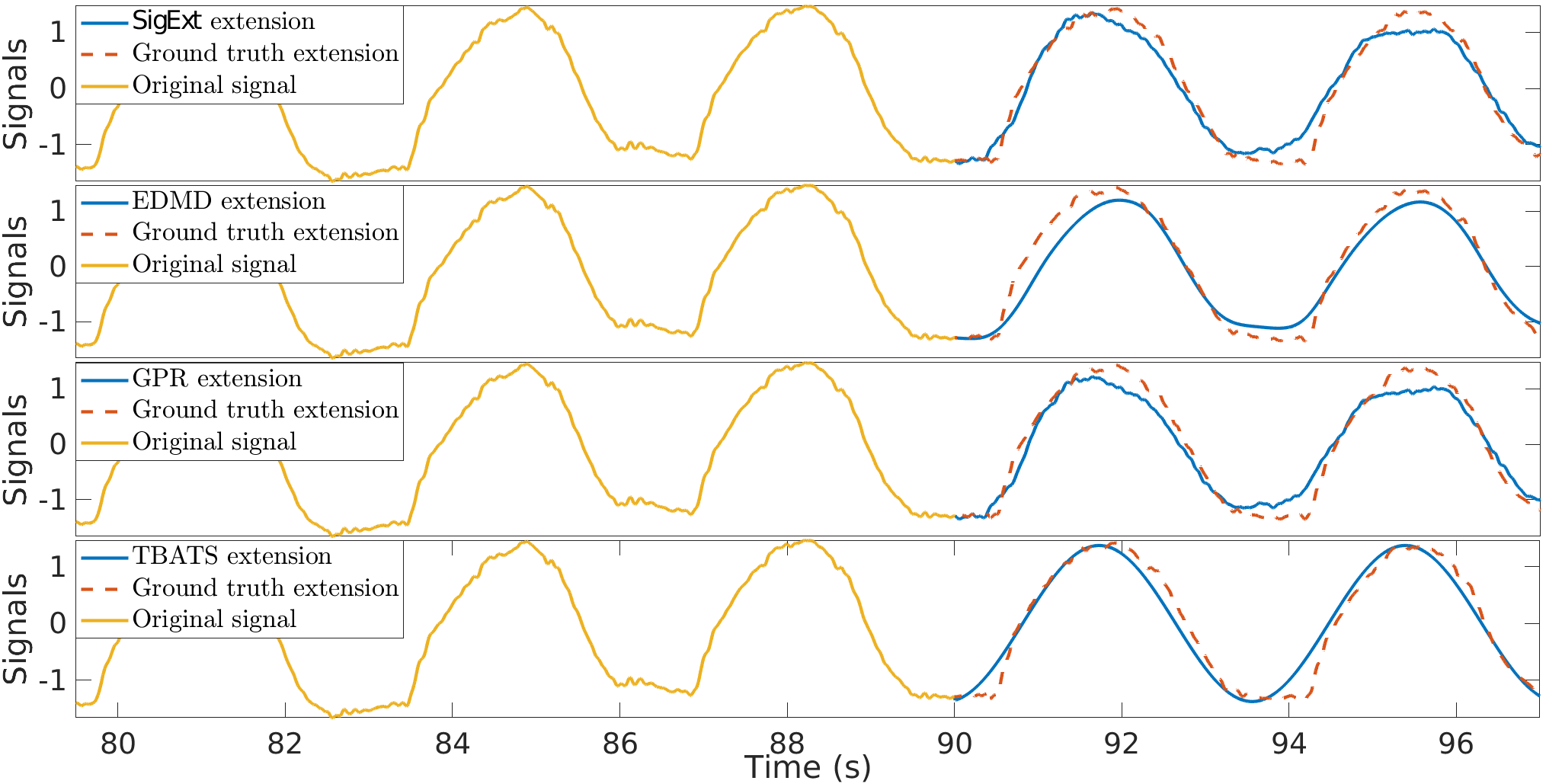}
\caption{Extended respiratory signal (blue) obtained by the {\sf SigExt} forecasting (first panel), the EDMD forecasting (second panel), the GPR forecasting (third panel), and the TBATS forecasting (fourth panel), superimposed with the ground truth signal (red-dashed line).}
\label{fig:tho}
\end{figure}

We first consider a thoracic movement signal recorded by the piezo-electrical sensor of length 6 hours and 20 minutes. The signal is sampled at $\fs=100$~Hz. A small portion of this signal is displayed in Fig.~\ref{fig:tho}.

\begin{figure}
\centering
\includegraphics[width=.48\textwidth]{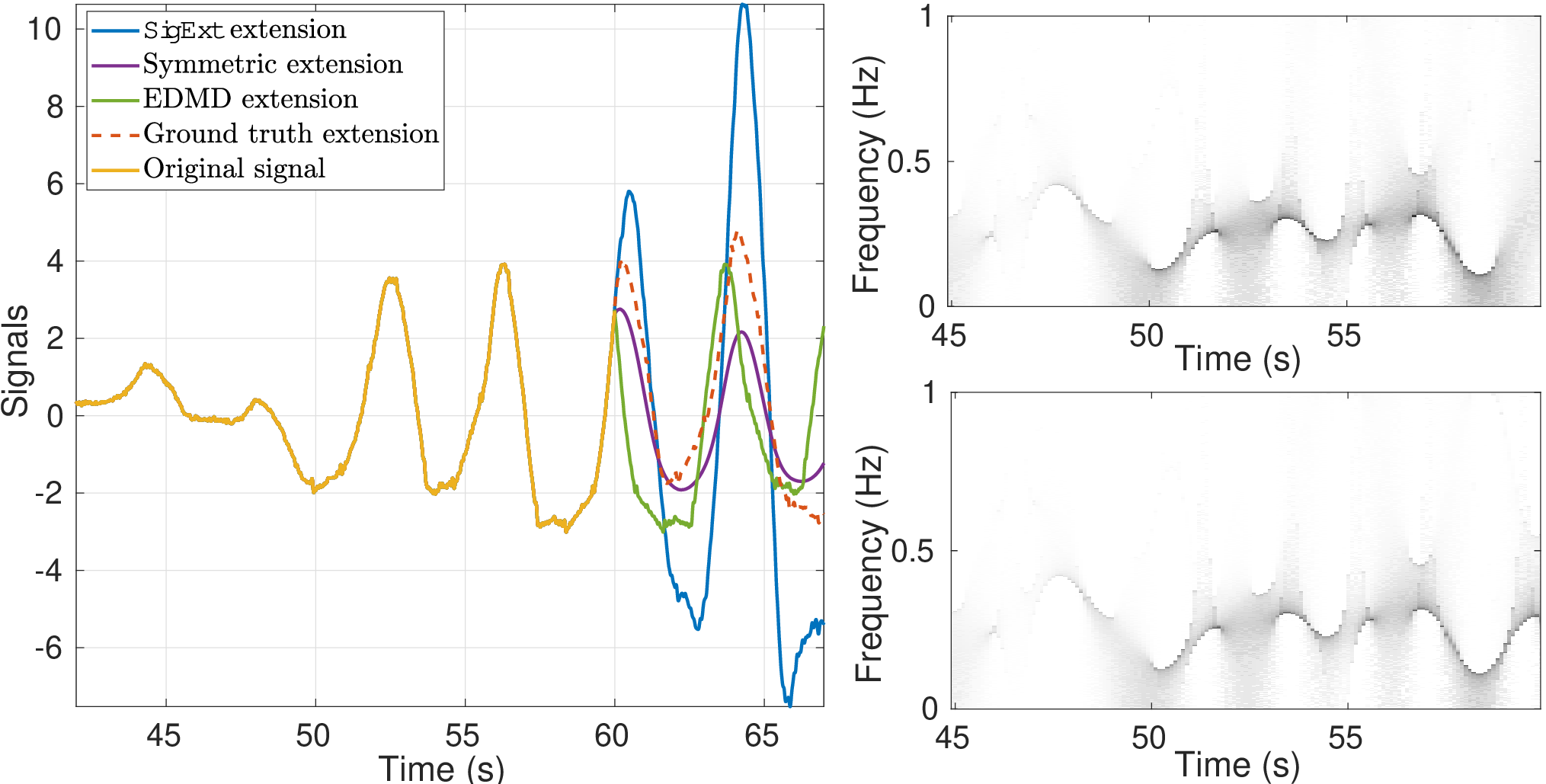}
\caption{Extensions of a segment of the respiratory signal (left) where {\sf SigExt} is outperformed by the EDMD and Symmetric extensions. Corresponding SST (top-right) and boundary-free SST obtained with {\sf SigExt} (bottom-right).}
\label{fig:THO.failure}
\end{figure}

From that long signal, we build a dataset of $379$ non\-overlapping signals of 60 seconds, \ie, $N=6000$. For each segment, we apply the forecasting methods introduced in Section~\ref{ssse:res.ahm}, including the {\sf SigExt} method detailed in Algorithm~\ref{alg:extension}. We forecast $7$-second-long extensions on each segment of the signal, corresponding to $L =700$. Thus, in order to catch the dynamical behavior, the size of the training signal $M$ is chosen so that $M=\lfloor 1.5L\rfloor$. As a result of Section~\ref{ssse:res.sine}, we take: $K=\lfloor2.5M\rfloor$. The extensions obtained on one of these subsignals are shown in Fig.~\ref{fig:tho}. The resulting MSEs of different methods are given in Table~\ref{tab:THO}. Note that the results are given in ``mean $\pm$ standard deviation'' format. The TBATS extension method is not implemented, as its excessive computing time makes it impossible to implement it on $6000$ segments within a reasonable period of time.
Note that the MSE of {\sf SigExt} is on average higher than the MSE of EDMD or the symmetric extension, with a huge standard deviation. This variability is caused by the presence of few segments contaminated by artifacts so that it is unpredictable via a too simple dynamical model like~\eqref{eq:dyn.model}. The left of Fig.~\ref{fig:THO.failure} illustrates one of these outliers, where {\sf SigExt} fails to catch the fast varying dynamic of the instantaneous amplitude to satisfactorily forecast the signal. The EDMD and symmetric extensions are more robust to those situations, as shown in this example, and in Table~\ref{tab:THO}. Nevertheless, {\sf SigExt} provides a sufficiently relevant extension to give TF representations sparingly affected by boundary effects. On the right of Fig.~\ref{fig:THO.failure}, we display a comparison between the right boundary of SST of the same segment of signal (top-right), and its boundary-free SST obtained after the {\sf SigExt} forecasting (bottom-right). The extension of the instantaneous frequency visible on the right side of the image illustrates the reduction of boundary effects produced despite an inaccurate signal forecasting.

We then apply {\sf BoundEffRed} for diverse TF representations: STFT, SST, RS, as well as \textit{Concentration of Frequency and Time} (ConceFT), a generalized multitaper SST-based representation introduced in~\cite{Daubechies16conceft}.  We mention that unlike other representations, the standard RS is not causal and rigorously requires knowing the STFT on the whole time-frequency plane before making any reassignment. To enable real-time implementation of RS, we use here a real-time version of RS proposed in~\cite{Lin17conceft}.

We set the extension length $L$ accordingly to the window length used by the TF analysis tool. For instance, the window length used to evaluate the STFT is of $1400$ samples. To prevent the STFT from being sensitive to the boundaries, we set $L=700$. In this way, the evaluation of the spectral content of the signal near its boundary is not limited by a lack of information in the interval delimited by the window support. From now on, all results are given for $L$ equal to the half of the width of the window used in the TF transform.
In Table~\ref{tab:THO}, we give the averaged performance index~\eqref{eq:index.perf} by evaluating the whole TF representations (including boundaries). Even though {\sf SigExt} performs moderately well in the sense of MSE, the boundary effects are dramatically reduced on the TF representations, and the averaged performance is in the same order of that given by EDMD or GPR. A t-test is performed to compare the performance index of {\sf SigExt} with those of other methods. Under the null hypothesis that the means are equal, with a $5\%$ significance level, the t-tests show no statistically significant difference between {\sf SigExt} and EDMD or GPR, regardless of the representation considered.

\begin{table}
\centering
\caption{Respiratory Signal: Performance of the Extension Methods and the Associated Boundary-Free TF Representations}
\begin{tabular}{|c||c||c|c|c|c|}
  \hline
   \multirow{2}{38pt}{\centering Extension method} & \multirow{2}{29pt}{\centering MSE} & \multicolumn{4}{c|}{Performance index $D$ (mean $\pm$ SD)} \\
   \cline{3-6}
      & & STFT & SST & RS & ConceFT \\
   \hhline{|=#=#=|=|=|=|}
   \multirow{2}{38pt}{\centering {\sf SigExt}} & $0.292$ & $0.370$ &  $0.408$ & $0.866$ & $0.423$ \\
   & $\pm 4.438$ & $\pm 0.623$ &  $\pm 0.436$ & $\pm 0.879$ & $\pm 0.344$ \\
   \hline
   \multirow{2}{38pt}{\centering Symmetric} & $0.044$ & $1.162$ & $1.173$ & $1.022$ & $1.144$ \\
   & $\pm 0.111$ & $\pm 0.893$ &  $\pm 0.886$ & $\pm 0.281$ & $\pm 0.579$ \\
   \hline
   \multirow{2}{38pt}{\centering EDMD} & $0.026$ & $0.359$ &  $0.422$ & $0.828$ & $0.449$ \\
   & $\pm 0.112$ & $\pm 0.266$ &  $\pm 0.282$ & $\pm 0.248$ & $\pm 0.296$ \\
   \hline
   \multirow{2}{38pt}{\centering GPR} & $0.331$ & $0.391$ &  $0.411$ & $0.897$ & $0.430$ \\ 
   & $\pm 4.858$ & $\pm 0.853$ &  $\pm 0.406$ & $\pm 1.140$ & $\pm 0.364$ \\
   \hline
\end{tabular}
\label{tab:THO}
\end{table}

\subsubsection{Two-Component Respiratory Signal}
\begin{figure}
\centering
\includegraphics[width=.48\textwidth]{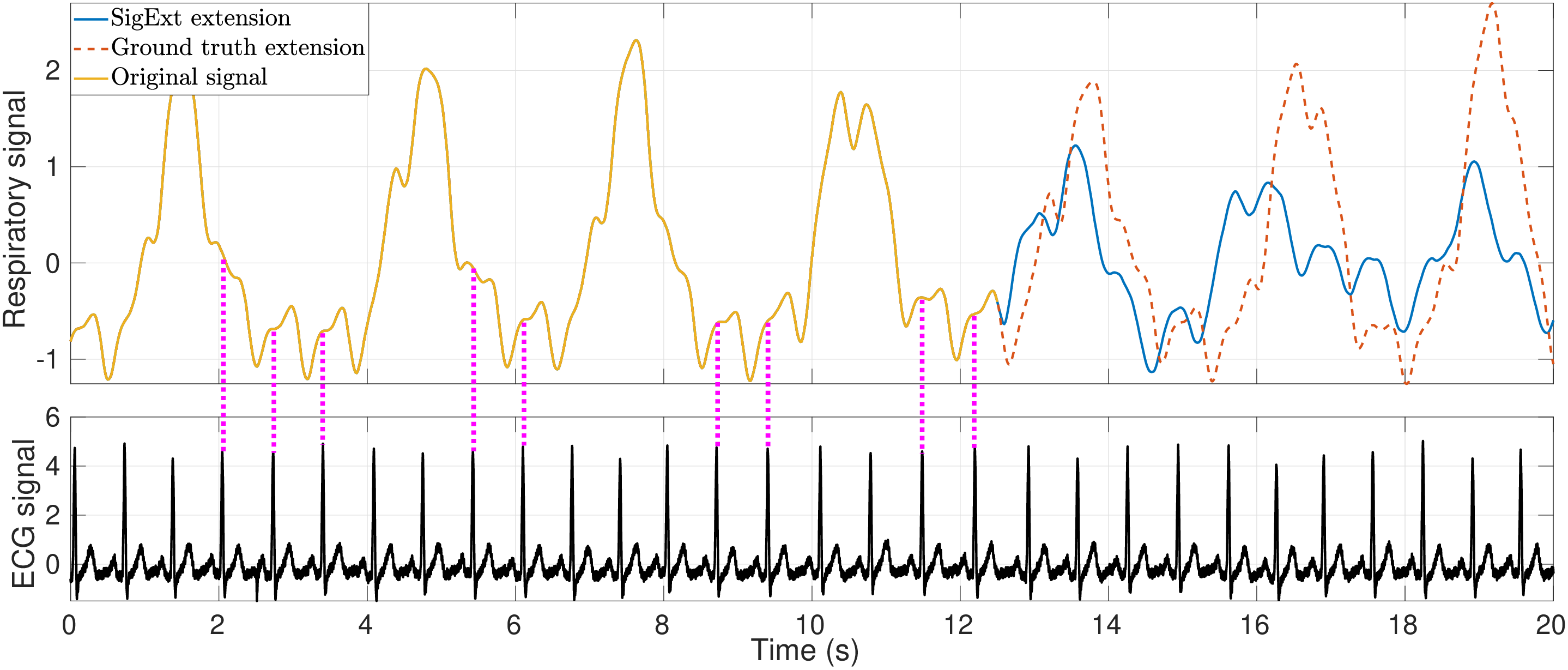}
\caption{Two-component respiratory signal (top, yellow), and the associated {\sf SigExt} extension (blue); the simultaneously recorded ECG signal is below. The vertical magenta dotted lines indicate the cardiac cycles in the respiratory signal.}
\label{fig:resp.2}
\end{figure}

Then, we consider a second respiratory signal that contains not only the respiratory cycle, but also a cardiac component known as the cardiogenic artifact~\cite{Smith94recognition}. The top of Fig.~\ref{fig:resp.2} shows an excerpt of the measured thoracic impedance-based respiratory signal, recorded during a bronchoscopic sedation procedure, along with the {\sf SigExt} extension. Below, the simultaneously recorded ECG signal is depicted. The magenta dotted lines emphasize the coincidences between the cardiogenic artifact of the respiratory signal and the QRS complexes of the ECG signal. Such cardiac information can be harvested and utilized when other first-line heart rate information resource is not available or broken~\cite{Lu19recycling}.

\begin{figure}
\centering
\includegraphics[width=.48\textwidth]{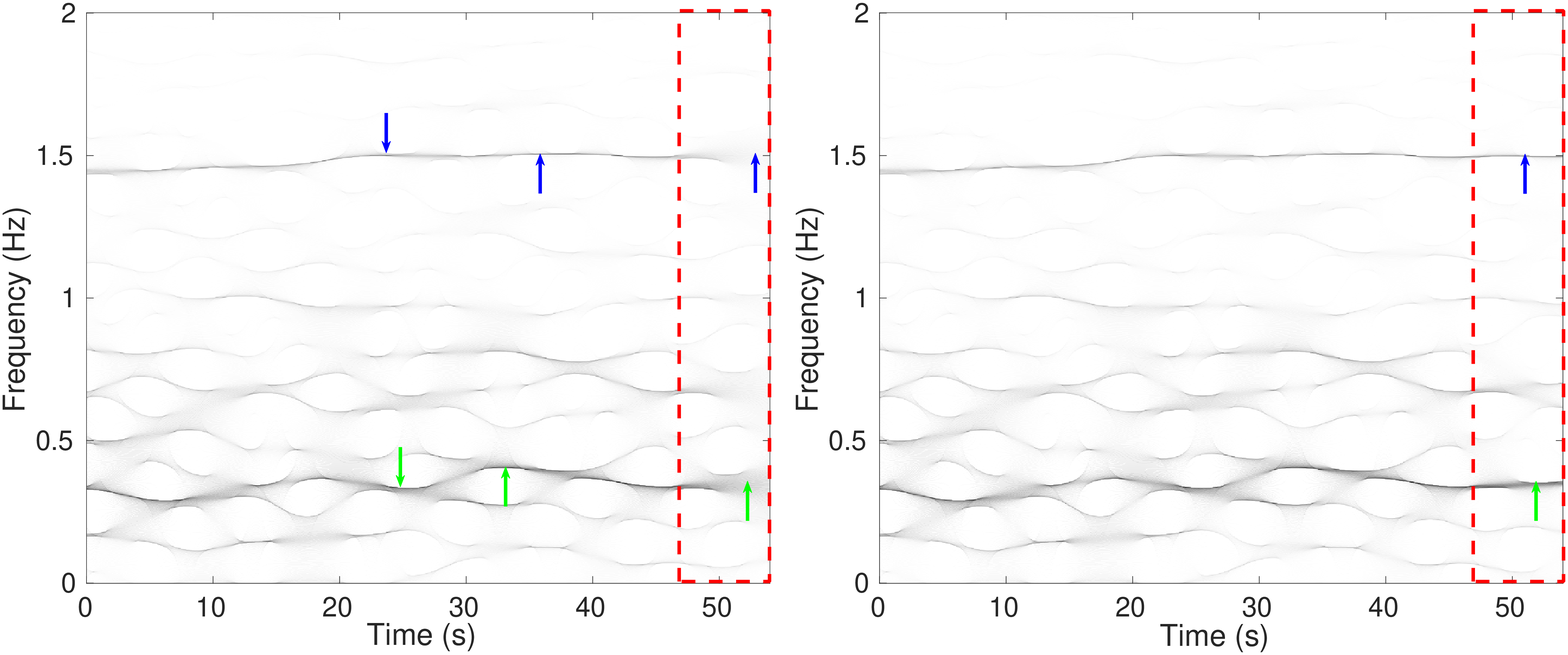}
\includegraphics[width=.48\textwidth]{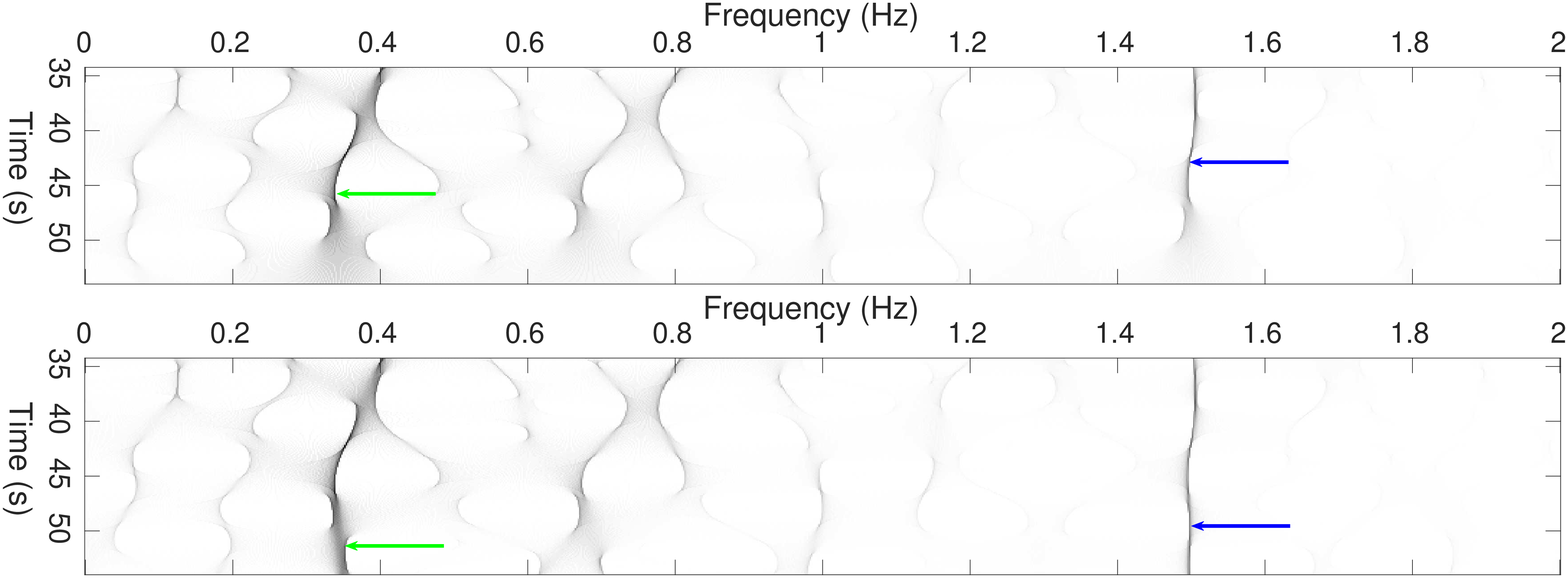}
\caption{Ordinary SST (top-left) and boundary-free SST (top-right) of a two-component respiratory signal. Middle and bottom: zoom on the areas delimited by the red dashed rectangles; these images are rotated 90 degrees clockwise. The window length for the SSTs is 20~seconds. The instantaneous frequency of the respiration is indicated by the green arrows, while the instantaneous frequency of the cardiogenic artifact is indicated by the blue arrows.}
\label{fig:resp.2.sst}
\end{figure}

The top of Fig.~\ref{fig:resp.2.sst} displays the ordinary SST and the boundary-free SST of this respiratory signal, sampled at $64$~Hz. This TF analysis brings out both components---the respiratory component, indicated by the green arrows, is located around the fundamental frequency $0.3$~Hz and its multiples, while the cardiac component, indicated by the blue arrows, is located around $1.5$~Hz. Note that, here, boundary effects are only reduced on the right-side boundary of the TF domain, delimited by the red dashed boxes. The middle of Fig.~\ref{fig:resp.2.sst} show a zoom near the right-boundary of the ordinary SST, while the bottom of Fig.~\ref{fig:resp.2.sst} respectively show the respective zoom on the boundary-free SST. In this area, the boundary-free SST makes it possible to disentangle the components contained in the signal. Moreover, the performance index of this representation takes the value $D=0.705$. This means that {\sf BoundEffRed} has reduced the right-side boundary effects by about $30\%$ with respect to the ordinary SST. This shows the ability of our algorithm to work on signals containing several nonstationary components.

\subsubsection{Photoplethysmogram}
\label{ssse:ppg}
\begin{figure}
\includegraphics[width=.48\textwidth]{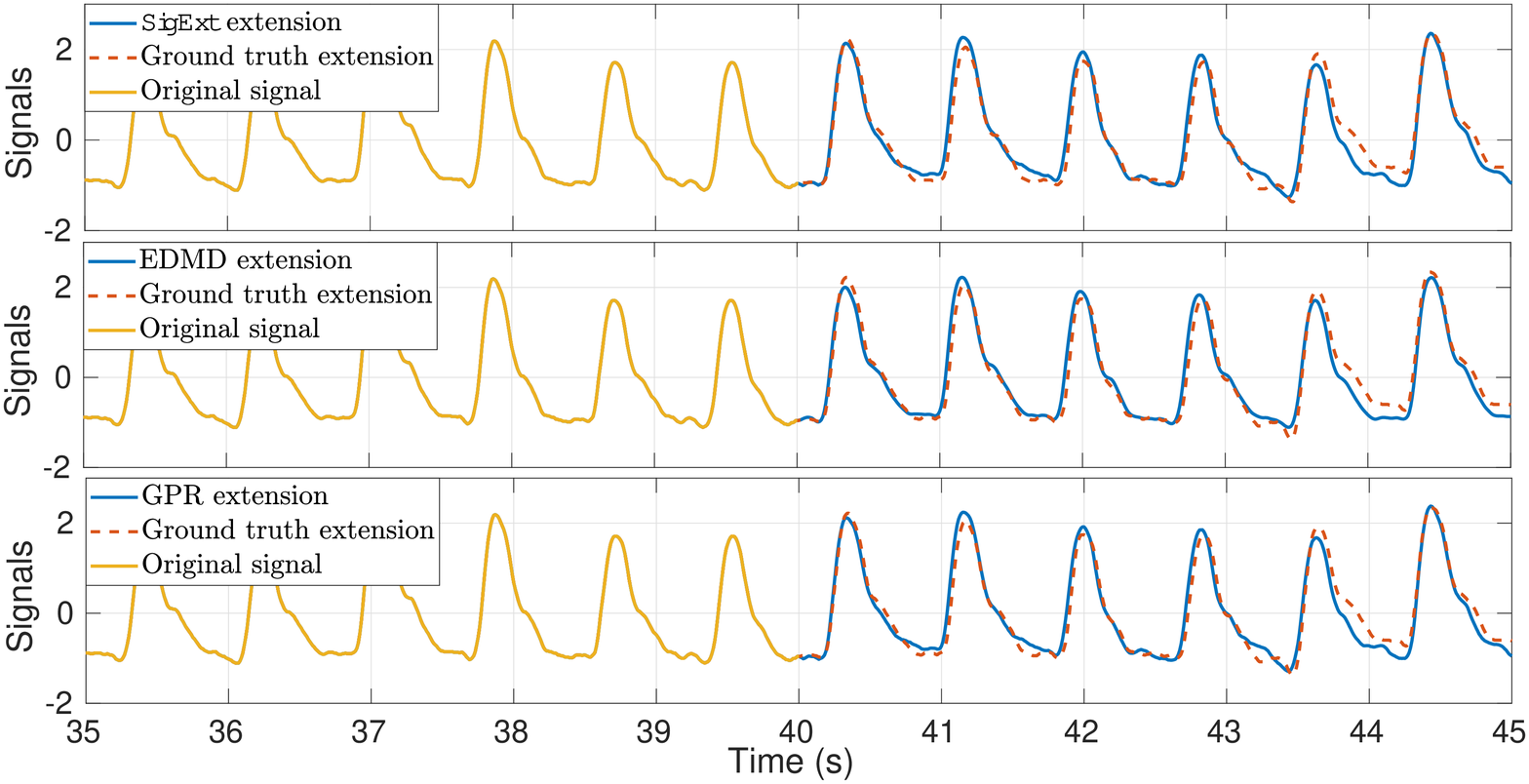}
\caption{Extended PPG signal (blue) obtained by the {\sf SigExt} forecasting (top), the EDMD forecasting (middle), and the GPR forecasting (bottom), superimposed with the ground truth signal (red dash).}
\label{fig:ppg}
\end{figure}

We consider a $640$-second photoplethysmogram (PPG) signal extracted from the Physionet dataset~\cite{Pimentel17toward, Goldberger00physiobank}, sampled at $\fs=125$~Hz. A portion of this signal is displayed in Fig.~\ref{fig:ppg}. The estimated 5-second extensions of this segment obtained by {\sf SigExt}, EDMD, and GPR forecastings  are superimposed to the ground-truth extension in Fig.~\ref{fig:ppg}.

We divide the signal into 32-second-long segments and apply {\sf BoundEffRed} to each of them. Table~\ref{tab:otd.ppg} shows the performance index $D$ of the boundary-free TF representations, averaged over the signals. For all the TF representations considered, the results show that {\sf SigExt} reduces boundary effects about as efficiently as the extensions given by EDMD or GPR. As for the respiratory signal, t-tests show no statistically significant difference between {\sf SigExt} and EDMD or GPR, regardless of the representation considered. For visual inspection, the TF representation of SST resulting from the {\sf BoundEffRed} strategy is shown in the bottom-right panel of Fig.~\ref{fig:ex.intro}, where SST is applied to the portion of PPG displayed in Fig.~\ref{fig:ppg}. It produces a significant improvement in the quality of the SST near boundaries. Indeed, the blurring visible when zooming on the right boundary of the SST has almost vanished. Real-time tracking of the instantaneous frequencies contained in the measured signal is therefore greatly facilitated.

\begin{table}
\centering
\caption{PPG Signal: Performance of the Boundary-Free TF Representations According to the Extension Method}
\begin{tabular}{|c||c||c|c|c|c|}
  \hline
   \multirow{2}{38pt}{\centering Extension method} & \multirow{2}{29pt}{\centering MSE} &\multicolumn{4}{c|}{Performance index $D$ (mean $\pm$ SD)} \\
   \cline{3-6}
      & & STFT & SST & ConceFT & RS\\
   \hhline{|=#=#=|=|=|=|}
   \multirow{2}{38pt}{\centering {\sf SigExt}} & $0.018$ & $0.280$ & $0.309$ & $0.367$ & $0.534$ \\
    & $\pm 0.014$ & $\pm 0.107$ & $\pm 0.112$ & $\pm 0.183$ & $\pm 0.160$ \\
   \hline
   \multirow{2}{38pt}{\centering Symmetric} & $0.037$ & $1.168$ & $1.209$ & $1.310$ & $0.983$ \\
   & $\pm 0.007$ & $\pm 0.390$ & $\pm 0.340$ & $\pm 0.140$ & $\pm 0.304$ \\
   \hline
   \multirow{2}{38pt}{\centering EDMD} & $0.012$ & $0.289$ & $0.319$ & $0.375$ & $0.503$ \\
   & $\pm 0.005$ & $\pm 0.126$ & $\pm 0.134$ & $\pm 0.163$ & $\pm 0.163$ \\
   \hline
   \multirow{2}{38pt}{\centering GPR} & $0.018$ & $0.276$ & $0.303$ & $0.361$ & $0.544$ \\
   & $\pm 0.013$ & $\pm 0.106$ & $\pm 0.110$ & $\pm 0.165$ & $\pm 0.157$ \\
   \hline
\end{tabular}
\label{tab:otd.ppg}
\end{table}

\begin{figure}
\centering
\includegraphics[width=.48\textwidth]{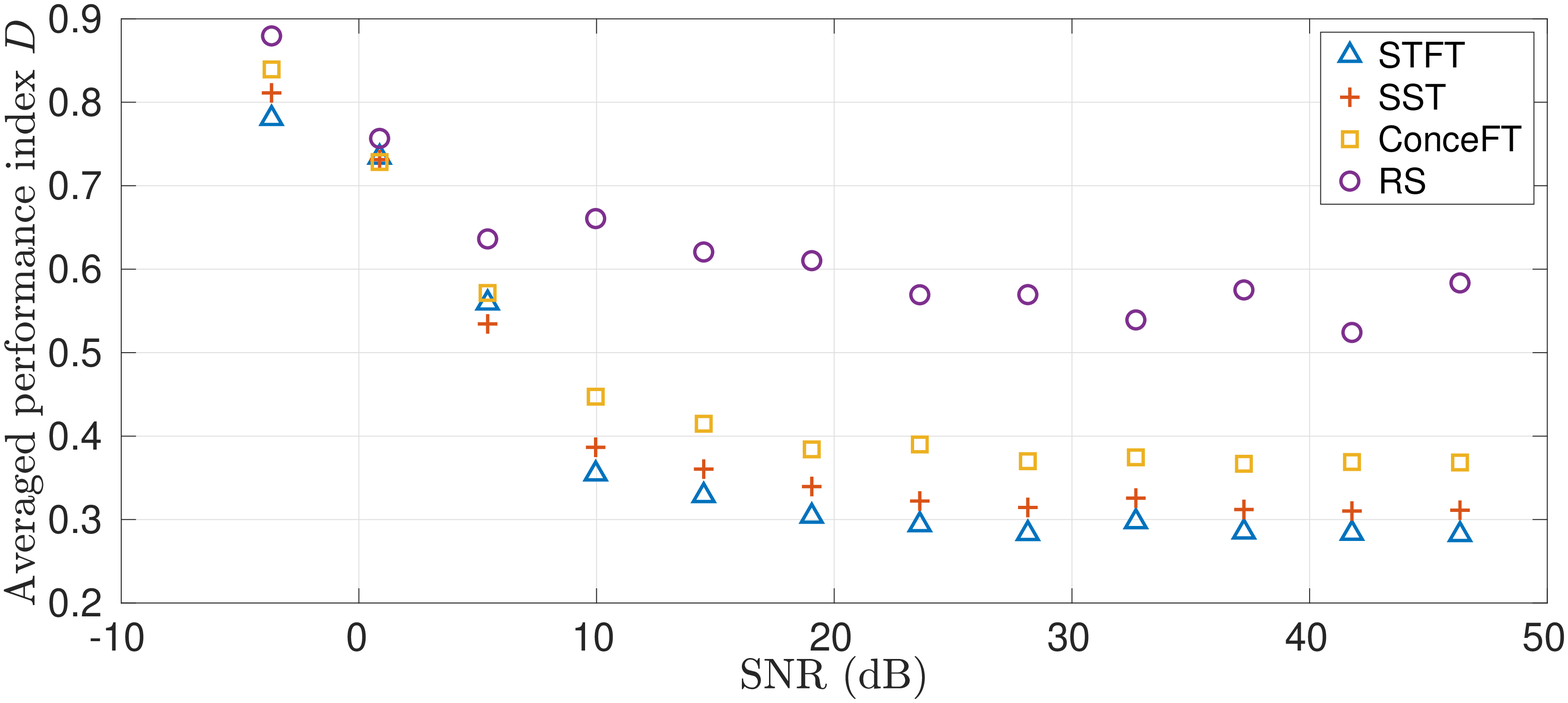}
\caption{Performance index of {\sf BoundEffRed} applied on a PPG, in function of the SNR.}
\label{fig:otd.noise}
\end{figure}

To further evaluate the influence of the noise level on the performance of {\sf BoundEffRed}, we artificially add Gaussian noise to the measured PPG signal. It is thus an additional noise to the measurement noise actually contained in the signal. Fig.~\ref{fig:otd.noise} shows the average performance index of {\sf BoundEffRed} for different values of the Signal-to-Noise Ratio (SNR). Whatever the representation considered, {\sf BoundEffRed} is relatively robust to noise, as long as the SNR remains above $10$~dB. Below this level, the reduction of boundary effects gradually deteriorates.

\subsection{Real-Time Implementation}
We simulate the real-time implementation of the SST from {\sf BoundEffRed} applied to the PPG described in Section~\ref{ssse:ppg}, subsampled by a factor of $2$. The test is performed on a 2-Core Intel Core i5 CPU running at $1.7$~GHz and $7.7$~GB of RAM. 

For this signal, sampled at $\fs=65.5$~Hz, the suitable window is $8$ seconds long. Therefore, the extension must be over $4$ seconds, \ie, $L=250$. The forecasting step in each iteration of {\sf BoundEffRed} then takes no more than $t_{\mathrm{forecast}}=46$~ms. Denote by $H\geq 1$ the hop size in samples, that is the number of samples between successive columns of the SST. Besides, since the support of the analysis window covers $2L$ samples, the update of the SST requires the calculation of $\lceil 2L/H \rceil$ new columns of the SST. In this example, the frequency dimension of the SST being $512$, the computational time of one column is $t_{\mathrm{SST}}=2.08$~ms, on average. A general rule to determine the acceptable values of $H$ for real-time implementation is verifying that the computational time to update the boundary-free TF representation is smaller than the lag between $H$ samples; that is,
\[
t_{\mathrm{forecast}} + \left\lceil \frac{L}H \right\rceil t_{\mathrm{SST}} < \frac{H}\fs\ . 
\]
In our example, taking $H\geq 8$~samples is sufficient to ensure the feasibility of real-time implementation. This value is reasonable because it allows a maximum overlap of $98.4\%$ of the window length. SST variations are therefore perceived gradually. Combined with the low latency of the algorithm, this allows for smooth monitoring of the real-time SST. A video clip illustrating this is available on GitHub, at \url{https://github.com/AdMeynard/BoundaryEffectsReduction/tree/master/Animations}. 